\pdfoutput=1
\documentclass[10pt]{article}
\usepackage[unicode]{hyperref}
\usepackage[utf8]{inputenc}
\usepackage[T1]{fontenc}
\usepackage[english]{babel}

\usepackage{amssymb, amsmath, amsthm}
\usepackage{mathtools}  
\usepackage{fullpage}
\usepackage{wrapfig}

\allowdisplaybreaks[3]

\usepackage[caption=false]{subfig}

\PassOptionsToPackage{monochrome}{xcolor}
\usepackage{tikz}
\usepackage{pgfplots}

\newcommand{\Kn}{\mathrm{Kn}}
\newcommand{\Ma}{\mathrm{Ma}}

\newcommand{\dd}{\mathrm{d}}
\newcommand{\pder}[2][]{\frac{\partial#1}{\partial#2}}
\newcommand{\pderdual}[2][]{\frac{\partial^2#1}{\partial#2^2}}
\newcommand{\pderder}[3][]{\frac{\partial^2#1}{\partial#2\partial#3}}
\newcommand{\Pder}[2][]{\partial#1/\partial#2}

\newcommand{\Pderder}[3][]{\partial^2#1/\partial#2\partial#3}
\newcommand{\dzeta}{\boldsymbol{\dd\zeta}}

\newcommand{\bzeta}{\boldsymbol{\zeta}}
\newcommand{\deltann}[2]{(\delta_{#1#2}-n_#1 n_#2)}

\newcommand{\bh}{\boldsymbol{h}}
\newcommand{\be}{\boldsymbol{e}}
\newcommand{\Nu}{\mathcal{N}}
\newcommand{\Mu}{\mathcal{M}}
\newcommand{\OO}[1]{O(#1)}
\newcommand{\oo}[1]{o(#1)}
\newcommand{\onwall}[1]{\left(#1\right)_0}
\newcommand{\Set}[2]{\{\,{#1}:{#2}\,\}}

\providecommand{\keywords}[1]{\textbf{\textit{Keywords:}} #1}
\title{Slow nonisothermal flows: numerical and asymptotic analysis of the Boltzmann equation}
\author{O.~A.~Rogozin\thanks{oleg.rogozin@phystech.edu, Moscow Institute of Physics and Technology,
    9 Institutskiy per., Dolgoprudny, Moscow Region, 141700, Russian Federation}}
\date{}

\begin{document}

\maketitle

\begin{abstract}
    \begin{flushright}
        {\it In memory of Osc\'{a}r Gavriilovich Friedl\'{a}nder (1939--2015)}
        \vspace{1em}
    \end{flushright}
    Slow flows of a slightly rarefied gas under high thermal stresses are considered.
    The correct fluid-dynamic description of this class of flows is based on the Kogan--Galkin--Friedlander equations,
    containing some non-Navier--Stokes terms in the momentum equation.
    Appropriate boundary conditions are determined from the asymptotic analysis of the Knudsen layer
    on the basis of the Boltzmann equation.
    Boundary conditions up to the second order of the Knudsen number are studied.
    Some two-dimensional examples are examined for their comparative analysis.
    The fluid-dynamic results are supported by numerical solution of the Boltzmann equation
    obtained by the Tcheremissine's projection-interpolation discrete-velocity method
    extended for nonuniform grids.
    The competition pattern between the first- and the second-order nonlinear thermal-stress flows
    has been obtained for the first time.
\end{abstract}


\section{Introduction}

From 1969 to 1974, Oscar Gavriilovich Friedlander (1939--2015) together with Vladlen Sergeevich Galkin (b.~1932)
under the guidance of Mikhail Naumovich Kogan (1925--2011) developed the theory of \emph{slow nonisothermal
slightly rarefied gas flows}~\cite{Kogan1970, Kogan1971, Friedlander1974, Galkin1974, Kogan1976}.
Slowness should be understood as the smallness of the Mach number (\(\Ma\ll1\)),
nonisothermality as the presence of significant temperature gradient in the gas (\(\nabla\log{T}\sim1\)),
and slightly rarefied means the smallness of the Knudsen number (\(\Kn\ll1\)).
The attempt to take into consideration the impact of thermal stress on the gas motion
through the analysis of the Burnett approximation was the main impetus of the mentioned works.
Despite the Burnett terms are of the second order of the Knudsen number,
they become comparable to the Newtonian viscous stresses in the case \(\Ma\sim\Kn\) (the Reynolds number is about unity).
Thus, the Navier--Stokes equations are incorrect to describe slow nonisothermal flows.

The correct set of fluid-dynamic-type equations was first obtained in~\cite{Kogan1970}.
There is no widely accepted term in the literature, so the author proposes to call them as
the \emph{Kogan--Galkin--Friedlander} or \emph{KGF equations}~\cite{Kogan1976}.
In the first papers~\cite{Kogan1970, Kogan1971}, these equations have been obtained the simplest way,
based on the Chapman--Enskog expansion, later from the Hilbert expansion~\cite{Galkin1974}.
The most general formulation of the time-dependent KGF equations for the mixture of gases can be found in~\cite{Galkin2015}.

In addition to the viscosity and thermal conductivity coefficients, the KGF equations contain
some \emph{thermal-stress transport coefficients}.
For some molecular potentials they were first calculated using the Sonine (associated Laguerre)
polynomials~\cite{Burnett1935, Chapman1970}.
For a hard-sphere gas, more accurate values were computed using the
direct numerical solution of corresponding integral equations~\cite{Sone1996, Sone2002, Sone2007}.

The nonlinear nature of the thermal stresses in the KGF equations leads to the phenomenon of gas convection
under their action (\emph{nonlinear thermal-stress convection})~\cite{Kogan1971}.
This type of convection takes place in the absence of external forces and may occur between uniformly heated bodies.
Nonlinear thermal stresses have also an influence on the process of heat transfer~\cite{Friedlander1978}.
After many years of work, Oscar Friedlander and his group succeeded in confirming
the theory of slow nonisothermal flows experimentally~\cite{Friedlander1997, Friedlander2003}.
At the same time, thanks to the development of computer technology,
some applied problems were analyzed numerically using the KGF and kinetic equations:
on the basis of model equations~\cite{Alexandrov2002, Aoki2006, Alexandrov2008b, Alexandrov2011, Rykov2008}
and the direct simulation Monte Carlo (DSMC) method~\cite{Alexandrov2008a, Aoki2007}.

For small Knudsen numbers, and especially slow flows, the classical DSMC method is extremely expensive.
Many specialized stochastic schemes have been proposed in recent years~\cite{Wagner2005, Hadji2013}.
Deterministic methods for the Boltzmann equation are also very promising~\cite{Dimarco2014, Mieussens2014}.
However, the computational difficulties have not been overcome completely up to date.
For this reason, there are no numerical examples of the nonlinear thermal-stress flows
on the basis of the Boltzmann equation, which require high accuracy of the numerical method, in the literature.
In addition, these flows cannot be observed in one-dimensional problems.
A high-accuracy analysis of the Knudsen layer is also a cumbersome task~\cite{Takata2015second},
especially around curved surfaces~\cite{Takata2015curvature}.
Slow flows with a wide temperature range impose additional requirements on the grid in the velocity space:
the distribution function of both cold and hot gas should be approximated equally well.
It can be satisfied, when the grid has a large volume, but refines for small molecular velocities.
The extension of the Tcheremissine's projection-interpolation discrete-velocity
method~\cite{Tcheremissine1997, Tcheremissine2000, Tcheremissine1998, Tcheremissine2006}
for nonuniform grids~\cite{Rogozin2016} based on the methodology of multi-point projection~\cite{Dodulad2012}
can effectively solve the considered class of problems.

Asymptotic analysis of the Boltzmann equation for slow nonisothermal flows shows
that the steady-state heat-conduction equation does not correctly describe a rarefied gas
in the continuum limit (\(\Kn\to0\))~\cite{Bobylev1995}.
This fact is also confirmed by numerical analysis~\cite{Sone1996}.
It turns out that infinitesimal velocity field has a finite impact on the temperature field.
This asymptotic behavior has been called the \emph{ghost effect}~\cite{Sone2002, Sone2007}.
Some mathematical questions of existence and stability of solutions to the KGF equations
are discussed in~\cite{Levermore2012, Tan2016}.

So far, the KGF equations were solved numerically only with the thermal-creep boundary conditions.
They are natural for the asymptotic equations of the leading order; however,
the solution can be improved if some boundary conditions of the following order are exploited,
such as temperature and speed jumps, thermal-stress slip, as well as those second-order terms
that comprise the curvature of the boundary surface.
This is possible since the above mention conditions depend only on the leading-order solution and some of its derivatives.

\section{Basic equations}

First of all, turn to dimensionless variables.
Let \(L\) be the reference length, \(T^{(0)}\) and \(p^{(0)}\) be the reference temperature and pressure of the gas.
Then the macroscopic variables take the following form: \(TT^{(0)}\) is the temperature,
\(pp^{(0)}\) is the pressure, \(\rho p^{(0)}/RT^{(0)}\) is the density, \(v_i(2RT^{(0)})^{1/2}\) is the velocity.
The distribution function \(f(x_i,\zeta_i)(2p^{(0)})/(2RT^{(0)})^{5/2}\) is defined
in the physical \(x_iL\) and velocity \(\zeta_i(2RT^{(0)})^{1/2}\) spaces.
The specific gas constant \(R = k_B/m\), where \(k_B\) is the Boltzmann constant, \(m\) is the mass of a molecule.
The Knudsen number \(\Kn = \ell^{(0)}/L\) is defined using the reference length free path
\begin{equation}\label{eq:ell}
    \ell^{(0)} = \frac{k_B T^{(0)}}{\sqrt2\pi d_m^2 p^{(0)}},
\end{equation}
where \(d_m\) coincides with the diameter of hard-sphere molecules.

In the presence of external forces \(F_i (2RT^{(0)})/L\), the dimensionless time-independent Boltzmann equation is
\begin{equation}\label{eq:Boltzmann}
    \zeta_i\pder[f]{x_i} + F_i\pder[f]{\zeta_i} = \frac1k J(f,f),
\end{equation}
where the collision integral
\begin{equation}\label{eq:integral}
    J(f,g) = \frac12 \int(f'g'_*+g'f'_*-fg_*-gf_*)B\dd\Omega(\boldsymbol\alpha) \dzeta_*
\end{equation}
and \(k = \sqrt\pi\Kn/2\).
\(\Omega(\boldsymbol{\alpha})\) is the solid angle in the direction of the unit vector \(\boldsymbol\alpha\),
\(B\) is the functional of the intermolecular potential. For a hard-sphere gas,
\begin{equation}\label{eq:ci_kernel}
    B = \frac{|\alpha_i(\zeta_{i*}-\zeta_i)|}{4\sqrt{2\pi}}.
\end{equation}
The macroscopic variables are expressed through the moments of the distribution function:
\begin{equation}\label{eq:macro}
    \rho = \int f \dzeta, \quad
    v_i = \frac1{\rho} \int \zeta_i f \dzeta, \quad
    T = \frac{2}{3\rho}\int(\zeta_i-v_i)^2 f \dzeta, \quad
    p = \rho T.
\end{equation}

The boundary conditions of diffuse reflection are specified as follows:
\begin{equation}\label{eq:diffuse_bc}
    f\left(\zeta_i n_i > 0\right) =
        \frac{\sigma_B}{(\pi T_B)^{3/2}} \exp\left(-\frac{\zeta_i^2}{T_B}\right), \quad
    \sigma_B = -2\left(\frac{\pi}{T_B}\right)^{1/2} \int_{\zeta_i n_i < 0} \zeta_j n_j f\dzeta,
\end{equation}
where \(n_i\) is the unit vector normal to the boundary, directed into gas,
\(T_B\) and \(v_{Bi}\) are the boundary temperature and velocity.
For time-independent problems, it is assumed that \(v_{Bi}n_i = 0\).

\section{Asymptotic analysis}

This section summarizes the main results of the asymptotic theory of slow nonisothermal flows
based on the Hilbert expansion.
The notation introduced in~\cite{Sone2002, Sone2007} is used.
For simplicity, only the hard-sphere molecular model is considered.

For small \(k\), the solution of time-independent Boltzmann equation can be separated
into different length scales. Consider the following form:
\begin{equation}\label{eq:sum_solutions}
    f = f_H + f_K,
\end{equation}
where \(f_H\) is the fluid-dynamic part of the solution on the scale \(\OO{1}\),
\(f_K\) is the Knudsen-layer correction on the scale \(\OO{k}\).
This separation is clear when \(f_K\) decreases faster than any inverse power of the distance from the boundary.
Nonlinear perturbation theory provides \(f_H\) that depends only on the macroscopic variables and their derivatives,
but \(f_H\) does not possess enough degrees of freedom to satisfy the diffuse-reflection boundary condition~\eqref{eq:diffuse_bc}.
Numerical solution of the Knudsen-layer problem yields \(f_K\), as well as the boundary conditions for \(f_H\).

\subsection{Fluid-dynamic part of the solution}

The distribution function \(f_H\) and the macroscopic variables \(h_H = \rho_H, v_{iH}, T_H, \dots\)
can be expanded in a power series of \(k\):
\begin{equation}\label{eq:expansion}
    f_H = f_{H0} + f_{H1}k + f_{H2}k^2 + \cdots, \quad h_H = h_{H0} + h_{H1}k + h_{H2}k^2 + \cdots.
\end{equation}
We look for a solution of the Boltzmann equation under the assumptions that
\(\Pder[f_H]{x_i} = \OO{f_H}\) (the Hilbert expansion), \(v_{iH0} = 0\) (slow flows),
\(F_{iH0} = F_{iH1} = 0\) (a weak field of external forces).
Substituting~\eqref{eq:expansion} in the Boltzmann equation~\eqref{eq:Boltzmann} and collecting terms of the same order of \(k\),
we obtain a system of integro-differential equations, for which the following solvability conditions must be satisfied:
in the zeroth order
\begin{equation}
    \pder[p_{H0}]{x_i} = 0; \label{eq:asymptotic0_p}
\end{equation}
in the first order
\begin{gather}
    \pder{x_i}\left(\frac{u_{iH1}}{T_{H0}}\right) = 0, \label{eq:asymptotic1_u} \\
    \pder[p_{H1}]{x_i} = 0, \label{eq:asymptotic1_p} \\
    \pder[u_{iH1}]{x_i} = \frac{\gamma_2}2\pder{x_i}\left(\sqrt{T_{H0}}\pder[T_{H0}]{x_i}\right); \label{eq:asymptotic1_T}
\end{gather}
for \(p_{H1} = 0\), in the second order
\begin{gather}
    \pder{x_i}\left(\frac{u_{iH2}}{T_{H0}}\right)
        = \frac{u_{iH1}}{T_{H0}}\pder{x_i}\left(\frac{T_{H1}}{T_{H0}}\right), \label{eq:asymptotic2_u} \\
    \begin{aligned}
        \pder{x_j}\left(\frac{u_{iH1}u_{jH1}}{T_{H0}}\right)
        &-\frac{\gamma_1}2\pder{x_j}\left[\sqrt{T_{H0}}\left(
            \pder[u_{iH1}]{x_j} + \pder[u_{jH1}]{x_i} - \frac23\pder[u_{kH1}]{x_k}\delta_{ij}
        \right)\right] \\
        &- \frac{\bar{\gamma}_7}{T_{H0}}\pder[T_{H0}]{x_i}\pder[T_{H0}]{x_j}\left(\frac{u_{jH1}}{\gamma_2\sqrt{T_{H0}}} - \frac{1}4\pder[T_{H0}]{x_j}\right)
        = -\frac12\pder[p_{H2}^\dag]{x_i} + \frac{p_{H0}^2 F_{iH2}}{T_{H0}},
    \end{aligned} \label{eq:asymptotic2_p} \\
    \pder[u_{iH2}]{x_i} = \frac{\gamma_2}2\pderdual{x_i}\left(T_{H1}\sqrt{T_{H0}}\right). \label{eq:asymptotic2_T}
\end{gather}
The following notation is introduced here: \(u_{iH1} = p_{H0}v_{iH1}\), \(u_{iH2} = p_{H0}v_{iH2}\), and
\begin{equation}\label{eq:dag_pressure}
    p_{H2}^\dag = p_{H0} p_{H2}
        + \frac{2\gamma_3}{3}\pder{x_k}\left(T_{H0}\pder[T_{H0}]{x_k}\right)
        - \frac{\bar{\gamma}_7}{6}\left(\pder[T_{H0}]{x_k}\right)^2.
\end{equation}
Equations~\eqref{eq:asymptotic1_u},~\eqref{eq:asymptotic1_T},~\eqref{eq:asymptotic2_p}
for \(T_{H0}\), \(u_{iH1}\), \(p_{H2}^\dag\) are proposed
to be called \emph{Kogan--Galkin--Friedlander} or \emph{KGF equations}~\cite{Kogan1976}.
They contain a thermal-stress term, which is absent in the Navier--Stokes equations.
Comparing it with \(p_{H0}^2F_{iH2}/T_{H0}\), we find that
\begin{equation}\label{eq:gamma7_force}
    k^2\frac{\bar{\gamma}_7}{p_{H0}^2}\pder[T_{H0}]{x_i}\pder[T_{H0}]{x_j}\left(\frac{u_{jH1}}{\gamma_2\sqrt{T_{H0}}} - \frac{1}4\pder[T_{H0}]{x_j}\right).
\end{equation}
is the force acting on unit mass of the gas.
It occurs when the isothermal surfaces are not parallel, i.e.,
\begin{equation}\label{eq:nonparallel}
    e_{ijk}\pder[T_{H0}]{x_j}\pder{x_k}\left(\pder[T_{H0}]{x_l}\right)^2 \ne 0.
\end{equation}
The Levi-Civita symbol \(e_{ijk}\) is used in~\eqref{eq:nonparallel}.
The gas motion driven by this force is called now the \emph{nonlinear thermal-stress flow}~\cite{Sone2002,Sone2007}.

Note that \(p_{H2}^\dag\) is not included directly in the equation of state and is therefore determined up to a constant.
Since the term \(\partial{p_{H2}^\dag}/\partial{x_i}\) is included in the system as the pressure
in the Navier--Stokes equations for incompressible gas,
the corresponding numerical methods are used to solve the KGF equations.

The dimensionless transport coefficients for a hard-sphere gas:
\begin{alignat*}{2}\label{eq:gamma_coeffs}
    \gamma_1 &= 1.270042427, &\quad \gamma_2 &= 1.922284066, \\
    \gamma_3 &= 1.947906335, &\quad \bar{\gamma}_7 &= 1.758705.
\end{alignat*}
The first two of them correspond, respectively, to the viscosity \(\mu\) and the thermal conductivity \(\lambda\) of the gas,
\begin{equation}\label{eq:mu_lambda}
    \mu = \gamma_1\sqrt{T_{H0}} \frac{p^{(0)}L}{\sqrt{2RT^{(0)}}} k, \quad
    \lambda = \frac{5\gamma_2}{2}\sqrt{T_{H0}} \frac{p^{(0)}RL}{\sqrt{2RT^{(0)}}} k.
\end{equation}
The coefficient \(\gamma_3\) is included in the thermal-stress expressions
that create nonuniform pressure distribution in the gas, but not the driving force.
Since \(\bar{\gamma}_7\) is positive, nonlinear thermal-stress flow is opposite to the temperature gradient.

\subsection{Knudsen layer and boundary conditions}

The diffuse-reflection condition~\eqref{eq:diffuse_bc} can be satisfied,
if we assume that \(f_K\) decreases exponentially on the scale of the mean free path in the proximity of a boundary:
\begin{gather}
    k n_i\pder[f_K]{x_i} = \OO{f_K}, \quad k\to0, \label{eq:knlayer_sharp}\\
    f_K = \oo{\eta^{-m}}, \quad \eta\to\infty, \quad m\in\mathbb{N}. \label{eq:knlayer_exp}
\end{gather}
Here, the natural orthogonal Knudsen-layer variables are introduced:
\begin{equation}\label{eq:eta_definition}
    x_i = k\eta n_i(\chi_1,\chi_2) + x_{Bi}(\chi_1,\chi_2),
\end{equation}
where \(x_{Bi}\) is the boundary surface, \(\eta\) is the stretched coordinate along the normal vector \(n_i\),
\(\chi_1\) and \(\chi_2\) are coordinates within the surface \(\eta=const\).
Then \(f_K\) satisfies
\begin{equation}\label{eq:fK_equation}
    \zeta_in_i\pder[f_K]{\eta} = 2J(f_H,f_K) + J(f_K,f_K)
    - k\zeta_i\left(\pder[\chi_1]{x_i}\pder[f_K]{\chi_1} + \pder[\chi_2]{x_i}\pder[f_K]{\chi_2}\right).
\end{equation}

The fluid-dynamic part of the zeroth-order solution is Maxwellian
\begin{equation}\label{eq:fH0_solution}
    f_{H0} = \frac{\rho_{H0}}{(\pi T_{H0})^{3/2}}\exp\left(-\frac{\zeta_i^2}{T_{H0}}\right),
\end{equation}
which satisfies~\eqref{eq:diffuse_bc} if
\begin{equation}\label{eq:boundary_T0}
    T_{H0} = T_{B0};
\end{equation}
therefore, \(f_{K0} = 0\). Using the expansions
\begin{gather}
    f_K = f_{K1}k + f_{K2}k^2 + \cdots, \label{eq:knlayer_expansion}\\
    f_H = \onwall{f_{H0}} + \left[\onwall{f_{H1}} + \eta\onwall{\pder[f_{H0}]{x_i}}n_i \right]k + \cdots, \label{eq:Honwall_expansion}
\end{gather}
where \(\onwall{\cdots}\) denotes the value of on the boundary (\(\eta=0\)),
the equations for \(f_{K1}\) and \(f_{K2}\) are
\begin{gather}
    \zeta_in_i\pder[f_{K1}]{\eta} = 2J\left[\onwall{f_{H0}},f_{K1}\right], \label{eq:knlayer1}\\
    \begin{multlined}
        \zeta_in_i\pder[f_{K2}]{\eta} = 2J\left[\onwall{f_{H0}}, f_{K2}\right]
        - \zeta_i\left[\onwall{\pder[\chi_1]{x_i}}\pder[f_{K1}]{\chi_1} + \onwall{\pder[\chi_2]{x_i}}\pder[f_{K1}]{\chi_2}\right] \\
        + 2J\left[\onwall{f_{H1}}+\eta\onwall{\pder[f_{H0}]{x_i}}n_i,f_{K1}\right] + J(f_{K1},f_{K1}).
    \end{multlined}\label{eq:knlayer2}
\end{gather}
Under appropriate boundary conditions in the half space \(\eta>0\), there exists a unique solution of
the one-dimensional linearized (about Maxwellian \(f_{H0}\) on the boundary) Boltzmann equation~\eqref{eq:knlayer1}
if and only if the boundary values of \(T_{H1}\) and \(u_{jH1}\deltann{i}{j}\) take specific values~\cite{Maslova1982, Bardos1986}.
A similar statement holds for~\eqref{eq:knlayer2} and the boundary values \(T_{H2}\), \(u_{jH2}\deltann{i}{j}\).

The homogeneous equation~\eqref{eq:knlayer1} leads to the following boundary conditions and Knudsen-layer corrections:
\begin{gather}
    \frac1{\sqrt{T_{B0}}}\begin{bmatrix} (u_{jH1} - u_{Bj2}) \\ u_{jK1} \end{bmatrix} \deltann{i}{j} =
        - \onwall{\pder[T_{H0}]{x_j}} \deltann{i}{j}
        \begin{bmatrix} K_1 \\ \frac12 Y_1(\tilde\eta) \end{bmatrix}, \label{eq:boundary_u1t}\\
    \begin{bmatrix} u_{jH1} \\ u_{jK1} \end{bmatrix} n_j = 0, \label{eq:boundary_u1n}\\
    \frac{p_{H0}}{T_{B0}}\begin{bmatrix} T_{H1} - T_{B1} \\ T_{K1} \\ T_{B0}^2\rho_{K1} \end{bmatrix} =
        \onwall{\pder[T_{H0}]{x_j}} n_j
        \begin{bmatrix} d_1 \\ \Theta_1(\tilde\eta) \\ p_{H0}\Omega_1(\tilde\eta) \end{bmatrix}, \label{eq:boundary_T1}
\end{gather}
where \(\tilde\eta = \eta p_{H0}/T_{B0}\).
The coefficients \(d_1\) and \(K_1\) correspond to the temperature jump and thermal creep, respectively.
For a hard-sphere gas,~\cite{Ohwada1989creep, Ohwada1989jump, Takata2015}
\begin{equation}\label{eq:slip_coeffs}
    d_1 = 2.40014, \quad K_1 = -0.64642.
\end{equation}
Since \(K_1<0\), the direction of the thermal creep coincides with the direction of the temperature gradient of the boundary surface.
The functions \(Y_1(\eta)\), \(\Theta_1(\eta)\), \(\Omega_1(\eta)\) decrease exponentially with \(\eta\)
and are tabulated for a hard-sphere gas in~\cite{Ohwada1989creep, Ohwada1989jump, Sone2002, Sone2007, Takata2015}.

In contrast to~\eqref{eq:knlayer1}, equation~\eqref{eq:knlayer2} is inhomogeneous.
In the absence of the last two nonlinear terms, this problem is well studied,
as appears in the asymptotic analysis of the linearized Boltzmann equation,
and leads to the following boundary conditions and Knudsen-layer corrections:
\begin{gather}
    \begin{aligned}
        \frac1{\sqrt{T_{B0}}}
            \begin{bmatrix} (u_{jH2} - u_{Bj2}) \\ u_{jK2} \end{bmatrix}\deltann{i}{j} =
        &- \frac{\sqrt{T_{B0}}}{p_{H0}}\onwall{\pder[u_{jH1}]{x_k}} \deltann{i}{j}n_k
            \begin{bmatrix} k_0 \\ Y_0(\tilde\eta) \end{bmatrix} \\
        - \frac{T_{B0}}{p_{H0}}\onwall{\pderder[T_{H0}]{x_j}{x_k}} \deltann{i}{j}n_k
            \begin{bmatrix} a_4 \\ Y_{a4}(\tilde\eta) \end{bmatrix}
        &- \bar\kappa\frac{T_{B0}}{p_{H0}}\onwall{\pder[T_{H0}]{x_j}} \deltann{i}{j}
            \begin{bmatrix} a_5 \\ Y_{a5}(\tilde\eta) \end{bmatrix} \\
        - \kappa_{jk}\frac{T_{B0}}{p_{H0}}\onwall{\pder[T_{H0}]{x_k}} \deltann{i}{j}
            \begin{bmatrix} a_6 \\ Y_{a6}(\tilde\eta) \end{bmatrix}
        &- \pder[T_{B1}]{x_j} \deltann{i}{j}
            \begin{bmatrix} K_1 \\ \frac12 Y_1(\tilde\eta) \end{bmatrix},
    \end{aligned}\label{eq:boundary_u2t}\\
    \begin{aligned}
        \frac1{\sqrt{T_{B0}}}
            &\begin{bmatrix} (u_{jH2} - u_{Bj2}) \\ u_{jK2} \end{bmatrix} n_j = \\
        &- \frac{T_{B0}}{p_{H0}}\left[ \onwall{\pderder[T_{H0}]{x_i}{x_j}}n_i n_j
            + 2\bar\kappa\onwall{\pder[T_{H0}]{x_i}}n_i \right]
            \begin{bmatrix} \frac12\int_0^\infty Y_1(\eta_0)\dd\eta_0 \\
                \frac12\int_\infty^{\tilde\eta} Y_1(\eta_0)\dd\eta_0 \end{bmatrix},
    \end{aligned}\label{eq:boundary_u2n}\\
    \begin{aligned}
        \frac{p_{H0}}{T_{B0}}
            \begin{bmatrix} T_{H2} - T_{B2} \\ T_{K2} \\ T_{B0}^2\rho_{K2} \end{bmatrix} =
        &\onwall{\pder[T_{H1}]{x_j}} n_j
            \begin{bmatrix} d_1 \\ \Theta_1(\tilde\eta) \\ p_{H0}\Omega_1(\tilde\eta) \end{bmatrix} \\
        + \frac{T_{B0}}{p_{H0}}\onwall{\pderder[T_{H0}]{x_i}{x_j}} n_i n_j
            \begin{bmatrix} d_3 \\ \Theta_3(\tilde\eta) \\ p_{H0}\Omega_3(\tilde\eta) \end{bmatrix}
        &+ \bar\kappa\frac{T_{B0}}{p_{H0}}\onwall{\pder[T_{H0}]{x_i}} n_i
            \begin{bmatrix} d_5 \\ \Theta_5(\tilde\eta) \\ p_{H0}\Omega_5(\tilde\eta) \end{bmatrix},
    \end{aligned}\label{eq:boundary_T2}
\end{gather}
where \(\bar\kappa/L = (\kappa_1+\kappa_2)/2L\) is the mean curvature of the boundary surface.
The principal curvatures, \(\kappa_1/L\) and \(\kappa_2/L\), become negative
when the corresponding center of curvature lies on the side of gas.
The dimensionless curvature tensor \(\kappa_{ij} = \kappa_1 l_i l_j + \kappa_2 m_i m_j\)
is expressed in terms of the direction cosines of the principal directions, \(l_i\) and \(m_i\).

The coefficient \(a_4\) corresponds to the second-order thermal-stress slip.
For a hard-sphere gas,~\cite{Ohwada1992, Takata2015}
\begin{equation}\label{eq:a4_coeff}
    a_4 = 0.0331.
\end{equation}
Since \(a_4>0\), there is a phenomenon of negative thermophoresis~\cite{Ohwada1992}.
The coefficients in front of \(\bar\kappa\) and \(\kappa_{ij}\)
are computed recently~\cite{Takata2015curvature, Takata2015}:
\begin{equation}\label{eq:curvature_coeffs}
    a_5 = 0.23353, \quad a_6 = -1.99878, \quad d_3 = 0.4993, \quad d_5 = 4.6180.
\end{equation}
Functions \(Y_{a4}(\eta)\), \(Y_{a5}(\eta)\), \(Y_{a6}(\eta)\), \(\Theta_3(\eta)\),
\(\Omega_3(\eta)\), \(\Theta_5(\eta)\), \(\Omega_5(\eta)\) also decrease exponentially with \(\eta\)
and are tabulated for a hard-sphere gas in~\cite{Ohwada1992, Sone2002, Sone2007, Takata2015curvature, Takata2015}.

The last two terms in~\eqref{eq:knlayer2} result in the additional nonlinear terms
in~\eqref{eq:boundary_u2t} and~\eqref{eq:boundary_T2}:
\begin{equation}\label{eq:boundary_nonlinear}
    \frac1{p_{H0}^2}\onwall{\pder[T_{H0}]{x_j}\deltann{i}{j}}\onwall{\pder[T_{H0}]{x_k}n_k}, \quad
    \frac1{p_{H0}^2}\onwall{\pder[T_{H0}]{x_i}n_i}^2;
\end{equation}
however, the complete solution of this inhomogeneous Knudsen-layer problem
for a hard-sphere gas is not presented in the literature.
For the model Krook--Welander equation~\cite{Krook1954, Welander1954},
numerical analysis of the second term (from~\eqref{eq:boundary_nonlinear}) is presented in~\cite{Sone1970}.

\subsection{Technique of using the next-order boundary conditions}

The next-order equations for \(T_{H1}\), \(v_{iH2}\), \(p_{H3}\) is cumbersome and
have not been obtained in the general form for an arbitrary molecular potential.
Therefore, numerical analysis of slow slightly rarefied flows is usually based on
the KGF equations~\eqref{eq:asymptotic1_u},~\eqref{eq:asymptotic1_T},~\eqref{eq:asymptotic2_p}
with the leading-order boundary conditions~\eqref{eq:boundary_T0},~\eqref{eq:boundary_u1t},~\eqref{eq:boundary_u1n}.
However, the asymptotic solution can be improved by introducing the known next-order boundary conditions.
For example, it is possible to calculate the temperature field \(T_H = T_{H0} + T_{H1}k + \OO{k^2}\) from the equation
\begin{equation}\label{eq:asymptotic_T}
    \frac1k\pder[u_{iH}]{x_i} = \frac{\gamma_2}2\pder{x_i}\left(\sqrt{T_H}\pder[T_H]{x_i}\right) + \OO{k^2},
\end{equation}
which is obtained from~\eqref{eq:asymptotic1_T} and~\eqref{eq:asymptotic2_T}, with the boundary condition
\begin{equation}\label{eq:boundary_T}
    T_H = T_B + d_1\frac{T_{B0}}{p_{H0}}\onwall{\pder[T_H]{x_j}}n_j k + \OO{k^2},
\end{equation}
which is obtained from~\eqref{eq:boundary_T0} and~\eqref{eq:boundary_T1}.
Since \(u_{iH2}\) is unknown, the temperature field \(T_H\) is calculated with the accuracy \(\OO{k}\),
but have the accuracy \(\OO{k^2}\) on the boundary.
The derivative of \(T_H\) is used in~\eqref{eq:boundary_T} instead of \(\Pder[T_{H0}]{x_j}\);
therefore, the temperature jump of the next-order boundary condition is taken into account.
Similarly, the velocity jump can be considered:
\begin{equation}\label{eq:boundary_u}
    u_{iH} = u_{Bi1}k - \left[ K_1\sqrt{T_{B0}}\onwall{\pder[T_{H0}]{x_j}}
        + k_0\frac{T_{B0}}{p_{H0}}\onwall{\pder[u_{jH}]{x_k}}n_k \right] \deltann{i}{j}k + \OO{k^2}.
\end{equation}
The terms from~\eqref{eq:boundary_u2t} and~\eqref{eq:boundary_T2}
that contain the second derivative of \(T_{H0}\) as well as \(\bar\kappa\) and \(\kappa_{ij}\)
can be included in the boundary conditions in the same way.
The boundary condition for the normal component of the velocity~\eqref{eq:boundary_u2n}
is incompatible with the equation~\eqref{eq:asymptotic1_u} and is therefore not used.

The fields \(T_H\) and \(u_{iH}\) obtained as described above describe the behavior of a rarefied gas
qualitatively better, because the additional boundary effects are taken into consideration.
Hence, one can also hope that they approximate the exact solution quantitatively better.

To calculate the second derivative of \(T_{H0}\) in the normal direction,
it is convenient to use the following transformation of~\eqref{eq:asymptotic1_T} and~\eqref{eq:boundary_u1t}:
\begin{equation}\label{eq:Tnn}
    \pderder[T_{H0}]{x_i}{x_j}n_i n_j + 2\bar\kappa\pder[T_{H0}]{x_i}n_i =
    - \pderdual[T_{H0}]{\chi_\alpha} - \frac1{2T_{H0}} \left[
        \left(\pder[T_{H0}]{x_i}n_i\right)^2 +
        \left(1+\frac{4K_1}{\gamma_2}\right) \left(\pder[T_{H0}]{\chi_\alpha}\right)^2
    \right],
\end{equation}
where each pair of repeated indices \(\alpha=1,2\) implies summation over them, and \(|\Pder[\chi_\alpha]{x_i}| = 1\).

\subsection{Computational implementation}

To solve the KGF equations~\eqref{eq:asymptotic1_u},~\eqref{eq:asymptotic1_T},~\eqref{eq:asymptotic2_p},
we employ the finite-volume method and the SIMPLE algorithm for pressure-velocity coupling~\cite{Aoki2007}.
The boundary conditions like~\eqref{eq:boundary_T} lead to the third-type boundary-value problem,
the solution of which requires additional measures to maintain stability of the numerical scheme.
The boundary temperature can become negative for large values of \(n_j\Pder[T_{H0}]{x_j}\).
To avoid this, it is usually sufficient to introduce a relaxation factor for the boundary conditions.
Moreover, the considered boundary-value problem has a solution in the limited range \(k<k_{\max}\).
The larger the temperature difference in the problem, the less \(k_{\max}\).
However, the solution of KGF equations is anyway unable to approximate the exact kinetic solution
adequately for larger \(k_{\max}\).

\subsection{Continuum limit}

In the classical fluid-dynamics, the Navier--Stokes equations (\(\bar{\gamma}_7=0\)
with the boundaries at rest (\(v_{Bi}=0\)) and the conditions without thermal creep (\(K_1=0\))
lead to the zero velocity field \(v_{iH1}=0\) and the heat-conduction equation
\begin{equation}\label{eq:heat_equation}
    \pder{x_i}\left(\sqrt{T_{H0}}\pder[T_{H0}]{x_i}\right) = 0.
\end{equation}
In the general case, the correct temperature distribution in the continuum limit (\(k\to0\))
is obtained from the KGF equations with appropriate boundary conditions.
It will coincide with the solution of~\eqref{eq:heat_equation} only for a narrow class of problems.

In the continuum world (\(k=0\)), there are no quantities like \(u_{iH1}\) and \(p^\dag_{H2}\);
nevertheless, infinitesimal velocity field \(v_i\) produce a finite effect on \(T\).
Such asymptotic behavior is called the ghost effect~\cite{Sone2002, Sone2007}.

\section{Method of solving the Boltzmann equation}

In the present work, the Boltzmann equation, written in such a dimensionless form
that the mean free path is the reference length
\begin{equation}\label{eq:Boltzmann_equation2}
    \pder[f]{t} + \zeta_i\pder[f]{x_i} = J(f),
\end{equation}
is solved numerically using the second-order symmetric splitting into the transport equation
\begin{equation}\label{eq:split_transport}
    \pder[f]{t} + \zeta_i\pder[f]{x_i} = 0,
\end{equation}
for which the finite-volume method with an explicit second-order TVD-scheme is employed,
and into the space-homogeneous Boltzmann equation
\begin{equation}\label{eq:split_collisions}
    \pder[f]{t} = J(f),
\end{equation}
for which the projection-interpolation discrete-velocity method
for non-uniform velocity grids is employed.
The brief description of the latter one is presented below.

\subsection{Discretization of the velocity space}

Let regular velocity grid \(\mathcal{V} = \Set{\zeta_\gamma}{\gamma\in\Gamma}\) is constructed in such a way
that the cubature over the molecular velocity space \(\bzeta\) is expressed as a weighted sum
\begin{equation}\label{eq:bzeta_cubature}
    \int F(\bzeta) \dzeta \approx \sum_{\gamma\in\Gamma} F_\gamma w_\gamma =
        \sum_{\gamma\in\Gamma} \hat{F_\gamma}, \quad
    \sum_{\gamma\in\Gamma} w_\gamma = V_\Gamma, \quad
    F_\gamma = F(\bzeta_\gamma),
\end{equation}
where \(F(\bzeta)\) is an arbitrary integrable function, \(V_\Gamma\) is the total volume of the velocity grid,
\(\Gamma\) is some index set.
Then, the eight-dimensional cubature formula in the space \((\boldsymbol{\omega},\bzeta,\bzeta_*)\) can be written as
\begin{equation}\label{eq:nu_cubature}
    \int F(\boldsymbol{\omega},\bzeta,\bzeta_*) \dd\Omega(\boldsymbol{\omega})\dzeta\dzeta_* \approx
        \frac{4\pi V_\Gamma^2}{ \sum_{\nu\in\Nu} w_{\alpha_\nu}w_{\beta_\nu} }
        \sum_{\nu\in\Nu} F(\boldsymbol{\omega}_\nu,\bzeta_{\alpha_\nu},\bzeta_{\beta_\nu}) w_{\alpha_\nu}w_{\beta_\nu},
\end{equation}
where \(F(\boldsymbol{\omega},\bzeta,\bzeta_*)\) is also an arbitrary integrable function.
\(\alpha_\nu\in\Gamma\), \(\beta_\nu\in\Gamma\),
and \(\boldsymbol{\omega}_\nu\in S^2 = \Set{\boldsymbol{\omega}\in\mathbb{R}^3}{|\boldsymbol{\omega}| = 1}\)
are obtained from some equal-weight cubature rule,
\(\Nu\subset\mathbb{N}\) is its index set.
Note that the numerical integration in~\eqref{eq:nu_cubature} is carried out over
the discrete spectrum of \((\bzeta,\bzeta_*)\) and continuous spectrum of \(\boldsymbol{\omega}\).

The collision integral written in the symmetrized form
\begin{equation}\label{eq:symm_ci}
    J(f_\gamma) = \frac14\int \left[
        \delta(\bzeta-\bzeta_\gamma) + \delta(\bzeta_*-\bzeta_\gamma)
        - \delta(\bzeta'-\bzeta_\gamma) - \delta(\bzeta'_*-\bzeta_\gamma)\right]
        (f'f'_* - ff_*)B \dd\Omega(\boldsymbol{\omega}) \dzeta\dzeta_*,
\end{equation}
where \(\delta(\bzeta)\) is the Dirac delta function in \(\mathbb{R}^3\),
has the following discrete analogue:
\begin{equation}\label{eq:discrete_symm_ci}
    \hat{J}_\gamma(\hat{f}_\gamma) =
        \frac{\pi V_\Gamma^2}{\sum_{\nu\in\Nu} w_{\alpha_\nu}w_{\beta_\nu}}
        \sum_{\nu\in\Nu} \left(
            \delta_{\alpha_\nu\gamma} + \delta_{\beta_\nu\gamma}
            - \delta_{\alpha'_\nu\gamma} - \delta_{\beta'_\nu\gamma}
        \right)\left(
            \frac{w_{\alpha_\nu}w_{\beta_\nu}}{w_{\alpha'_\nu}w_{\beta'_\nu}}
            \hat{f}_{\alpha'_\nu}\hat{f}_{\beta'_\nu} - \hat{f}_{\alpha_\nu}\hat{f}_{\beta_\nu}
        \right)B_\nu,
\end{equation}
where \(\delta_{\varsigma\gamma}\) is the Kronecker delta.
In the general case, \(\bzeta_{\alpha'_\nu}\) and \(\bzeta_{\beta'_\nu}\) are not in \(\mathcal{V}\);
therefore, quantities \(\hat{f}_{\alpha'_\nu}\), \(\hat{f}_{\beta'_\nu}\), \(w_{\alpha_\nu}\), \(w_{\beta_\nu}\)
and functions \(\delta_{\alpha'_\nu\gamma}\), \(\delta_{\beta'_\nu\gamma}\) have to be defined in some way.

The Maxwell distribution is approximated as follows:
\begin{equation}\label{eq:discrete_Maxwellian}
    \hat{f}_{M\gamma} = \rho\left[\sum_{\varsigma\in\Gamma}w_\varsigma\exp
            \left(-\frac{(\bzeta_\varsigma - \boldsymbol{v})^2}{T}\right)
        \right]^{-1}
        w_\gamma\exp\left(-\frac{(\bzeta_\gamma - \boldsymbol{v})^2}{T}\right).
\end{equation}

\subsection{Projection-interpolation technique}

If the velocities after collision,
\(\bzeta_{\alpha'_\nu}\notin\mathcal{V}\) and \(\bzeta_{\beta'_\nu}\notin\mathcal{V}\),
are replaced with the nearest grid velocities,
\(\bzeta_{\lambda_\nu}\in\mathcal{V}\) and \(\bzeta_{\mu_\nu}\in\mathcal{V}\),
the discrete collision integral~\eqref{eq:discrete_symm_ci} is not strictly conservative,
and the discrete Maxwellian~\eqref{eq:discrete_Maxwellian} is not the equilibrium state.
To resolve these issues, two special procedures are applied in the projection-interpolation method.
First, \(\bzeta_{\alpha'_\nu}\) is projected to a set of grid velocities
\(\Set{\bzeta_{\lambda_\nu+s_a}}{a\in\Lambda}\subset\mathcal{V}\) in the following way:
\begin{equation}\label{eq:ci_projection}
    \delta_{\alpha'_\nu\gamma} = \sum_{a\in\Lambda} r_{\lambda_\nu,a}\delta_{\lambda_\nu+s_a,\gamma},
\end{equation}
where the index set \(\Lambda = \Set{a}{r_{\lambda_\nu,a}\neq0}\subset\mathbb{Z}\).
The set of displacement rules \(\mathcal{S} = \Set{s_a}{a\in\Lambda}\)
is called the \emph{projection stencil}.

The expression~\eqref{eq:ci_projection} can be formally regarded as an approximate solution of the equation
\(\phi=\delta(\bzeta'-\bzeta_\gamma)\) in the space of delta functions
\(\Set{\delta(\bzeta-\bzeta_\gamma)}{\bzeta_\gamma\in\Nu}\)
by the projection Petrov--Galerkin method onto a linear span of functions \(\psi_s(\bzeta)\):
\begin{equation}\label{eq:Petrov-Galerkin}
    \int \psi_s(\bzeta_\gamma) \left( \delta(\bzeta'-\bzeta_\gamma)
        - \sum_{a\in\Lambda} r_{\lambda_\nu,a} \delta(\bzeta_{\lambda_\nu+s_a}-\bzeta_\gamma) \right) \dzeta_\gamma = 0.
\end{equation}
If the set \(\{\psi_s\}\) contains all the collision invariants, for example,
\begin{equation}\label{eq:collision_invariants}
    \psi_0 = 1, \quad \psi_i = \zeta_i, \quad \psi_4 = \zeta_i^2,
\end{equation}
then each cubature point (term in~\eqref{eq:discrete_symm_ci}) ensures the conservation of mass, momentum and kinetic energy
for the found \textit{projective velocities} \(\bzeta_{\lambda_\nu+s_a}\) and \textit{projection weights} \(r_{\lambda_\nu a}\).

Second, to satisfy
\begin{equation}\label{eq:strict_interpolation}
    J_\gamma\left(\hat{f}_{M\gamma}\right) = 0,
\end{equation}
\(\hat{f}_{\alpha'_\nu}\) is interpolated in the following way:
\begin{equation}\label{eq:ci_interpolation}
    \frac{\hat{f}_{\alpha'_\nu}}{w_{\alpha'_\nu}} = \prod_{a\in\Lambda}
        \left(\frac{\hat{f}_{\lambda_\nu+s_a}}{w_{\lambda_\nu+s_a}} \right)^{r_{\lambda_\nu,a}}.
\end{equation}
This type of interpolation has a large computational cost,
but is strictly required for low-noise analysis of slow flows,
when the distribution function is close to the Maxwellian.
In practice, the exponentiation can be performed with an error of about \(10^{-5}\),
allowing several times to speed up the calculations.
In addition,~\eqref{eq:ci_interpolation} ensures that
the Boltzmann H-theorem holds in the discrete form~\cite{Dodulad2013}.
For \(\bzeta_{\beta'_\nu}\) and \(\hat{f}_{\beta'_\nu}\), all formulas are similar.

\subsection{Solution of the Cauchy problem}

Now turn to the space-homogeneous Boltzmann equation~\eqref{eq:split_collisions}.
Let \(f_\gamma^n\) denotes an approximate solution of~\eqref{eq:split_collisions}
for velocity \(\bzeta_\gamma\) (\(\gamma\in\Gamma\)) at time \(t_n\) (\(n\in\mathbb{N}\)).
Rewriting~\eqref{eq:discrete_symm_ci} as
\begin{equation}\label{eq:discrete_short_ci}
    \hat{J}_\gamma^n\left(\hat{f}_\gamma^n\right) =
        \sum_{\nu=1}^N \hat{\Delta}_\gamma^{n+(\nu-1)/N} \left(\hat{f}_\gamma^n\right), \quad
    N=|\Nu|,
\end{equation}
where \(\hat{\Delta}_\gamma^{n+(\nu-1)/N}\) is \(\nu\in\Nu_n\) term of sum~\eqref{eq:discrete_symm_ci},
one can apply the first-order explicit Euler method in fractional steps:
\begin{equation}\label{eq:time_scheme}
    \hat{f}_\gamma^{n+\nu/N} = \hat{f}_\gamma^{n+(\nu-1)/N} + \tau \hat{\Delta}_\gamma^{n+(\nu-1)/N}
    \left(\hat{f}_\gamma^{n+(\nu-1)/N}\right),
\end{equation}
where \(\tau = t_{n+1} - t_n\) is the time step.
Scheme~\eqref{eq:time_scheme} has a convergence rate \(\OO{\tau|\Gamma|/|\Nu|}\)
if all of the discrete velocities \(\bzeta_\gamma\) are distributed uniformly
in the sequences \((\alpha_\nu)_{\nu=1}^N\) and \((\beta_\nu)_{\nu=1}^N\).
This can be accomplished by a random permutation of the cubature sequence.
For \(|\Gamma|/|\Nu| = \OO{\tau}\), the second-order accuracy is achieved.

The Korobov lattice rule~\cite{Korobov1959, Sloan1994} is used as
an equal-weight cubature rule in~\eqref{eq:discrete_symm_ci}.
The integration lattice is randomly shifted every time step
to obtain a sequence of sets of cubature points \((\Nu_n)_{n\in\mathbb{N}}\).

\subsection{Preservation of positivity}

Scheme~\eqref{eq:time_scheme} allows negative values of the distribution function
and loses stability in the presence of them.
To ensure its positivity, it is enough to require
\begin{equation}\label{eq:positive_f}
    \hat{f}_\gamma^{n+(j-1)/N} + \frac{\tau}N \hat{\Delta}_\gamma^{n+(j-1)/N} > 0
\end{equation}
for all \(\gamma\in\Gamma\) and \(\nu\in\Nu_n\).
For \(\gamma = \alpha_\nu\),
\begin{equation}\label{eq:positive_f_alpha}
    \hat{f}_{\alpha_\nu} - \frac{A}{N}\hat{f}_{\alpha_\nu}\hat{f}_{\beta_\nu} > 0, \quad
    A = \frac{\pi\tau V_\Gamma^2 N B_{\max}}{\sum_{\nu\in\Nu} w_{\alpha_\nu}w_{\beta_\nu}}
\end{equation}
or
\begin{equation}\label{eq:positive_f_alpha2}
    N > A \hat{f}_{\max},
\end{equation}
where
\begin{equation}\label{eq:f_B_max}
    \hat{f}_{\max} = \max_{\gamma\in\Gamma} \hat{f}_\gamma, \quad
    B_{\max} = \max_{\substack{\gamma,\varsigma\in\Gamma\\\boldsymbol{\omega}\in S^2}}
        B(\boldsymbol{\omega}, \bzeta_{\gamma}, \bzeta_{\varsigma}) = \OO{\zeta_{\max}}, \quad
    \zeta_{\max} = \max_{\gamma\in\Gamma}|\bzeta_\gamma|.
\end{equation}
The same estimate holds for \(\gamma = \beta_\nu\).

The projection nodes \(\gamma = \lambda_\nu+s_a\) (and \(\gamma = \mu_\nu+s_a\))
are treated with interpolation~\eqref{eq:ci_interpolation}.
Additionally, assume \(r_{\lambda_\nu,a} \leq 1\).
For \(r_{\lambda_\nu,a} \geq 0\),
\begin{equation}\label{eq:positive_f_lambda2+}
    N > A \hat{f}_{\max} \epsilon_f^2 \epsilon_w^2,
\end{equation}
where
\begin{equation}\label{eq:epsilon_f}
    \epsilon_f = \max_{\substack{s_a,s_b\in\mathcal{S}\\\gamma\in\Gamma}} \frac{\hat{f}_{\gamma+s_a}}{\hat{f}_{\gamma+s_b}}, \quad
    \epsilon_w = \max_{\gamma,\varsigma\in\Gamma} \frac{w_\gamma}{w_\varsigma}.
\end{equation}
For a smooth distribution function, \(\epsilon_f\) is proportional to the maximum \emph{diameter} of the projection stencil
\begin{equation}\label{eq:stencil_diameter}
    R_\mathcal{S} = \max_{\substack{s_a,s_b\in\mathcal{S}\\\gamma\in\Gamma}}
        \left| \bzeta_{\gamma+s_a} - \bzeta_{\gamma+s_b} \right|.
\end{equation}
For \(r_{\lambda_\nu,a} < 0\),
\begin{equation}\label{eq:positive_f_lambda-}
    \hat{f}_{\lambda_\nu+s_a} + \frac{A}{N}r_{\lambda_\nu,a} \hat{f}_{\alpha_\nu}\hat{f}_{\beta_\nu} > 0.
\end{equation}
For an arbitrary distribution function, we obtain a very expensive estimate
\begin{equation}\label{eq:positive_f_lambda2-}
    N > A \hat{f}_{\max} \bar{r}_{\max} \max_{\gamma,\varsigma\in\Gamma}\frac{\hat{f_\gamma}}{\hat{f_\varsigma}}, \quad
    \bar{r}_{\max} = \max_{\gamma\in\Gamma,a\in\Lambda}( -r_{\gamma,a} ),
\end{equation}
but for a Maxwellian,
\begin{equation}\label{eq:positive_f_lambda2-maxw}
    N > A \hat{f}_{\max} \epsilon_f^2 \bar{r}_{\max}.
\end{equation}

In summary, to decrease \(N\) required for~\eqref{eq:positive_f},
one should construct a velocity grid that minimizes \(|\Gamma|\), \(\zeta_{\max}\), \(\epsilon_w\)
and a projection stencil that minimizes \(R_\mathcal{S}\), \(\bar{r}_{\max}\).
\(\epsilon_f\) can be lowered by means of grid refinement in regions of large gradients of the distribution function.

In practice, condition~\eqref{eq:positive_f} for all \(\nu\) requires a large computational cost.
In order to achieve an acceptable accuracy, it is sufficient to exclude terms
that violate~\eqref{eq:positive_f} from~\eqref{eq:time_scheme}.
In other words, the collision integral can be calculated as
\begin{equation}\label{eq:discrete_short_ci_discarded}
    \hat{J}_\gamma^n = \sum_{\nu\in\Nu\setminus\Mu} \hat{\Delta}_\gamma^{n+(\nu-1)/N},
\end{equation}
where \(\Mu\) is the set of cubature points excluded from \(\Nu\).
To avoid a significant error in this integration method,
it is necessary to control the contribution of excluded points to the collision integral.
For example, \(N\) can be adjusted so that
\begin{equation}\label{eq:excluded_contribution}
    \epsilon_J = \frac{\pi V_\Gamma^2}{\rho\sum_{\nu\in\Nu} w_{\alpha_\nu}w_{\beta_\nu}}
        \sum_{\nu\in\Mu} \left|
            \hat{f}_{\lambda_\nu}\hat{f}_{\mu_\nu} - \hat{f}_{\alpha_\nu}\hat{f}_{\beta_\nu}
        \right|B_\nu
\end{equation}
appears to be sufficiently small.
Interpolation~\eqref{eq:ci_interpolation} may lead to enormous values of \(\hat{f}_{\alpha'_\nu}\) when
one of \(\hat{f}_{\lambda_\nu+s_a}\) is extremely small and the corresponding weight \(r_{\lambda_\nu,a}\) is negative.
For this reason, we avoid interpolation in~\eqref{eq:excluded_contribution}.

\subsection{Projection stencils}

Hereinafter, the velocity grid is assumed to be rectangular in \(\mathbb{R}^3\);
therefore, it can be indexed by an integer vector, i.e., \(\Gamma = \Set{\gamma}{\gamma\in\mathbb{Z}^3}\).
A displacement rule can be also represented as an integer vector, i.e., \(\mathcal{S}\subset\mathbb{Z}^3\).
In this way, a sum of the indexes should be interpreted as a vector sum in \(\mathbb{Z}^3\).
The projection method is second-order accurate with respect to the step of a rectangular velocity grid~\cite{Anikin2012}.

Due to the symmetry of a uniform grid,
it is sufficient to use two projection nodes to ensure the conservative laws.
In the general case, five projection nodes are necessary for the existence of
a solution of~\eqref{eq:Petrov-Galerkin} for~\eqref{eq:collision_invariants}
The diameter of the stencil \(R_\mathcal{S}\) can be narrowed if seven projection nodes are employed.
When \(|\mathcal{S}|=n\), scheme~\eqref{eq:time_scheme} is called the \(n\)-\emph{point scheme}.

The \emph{2-point scheme} is based on the symmetric projection
\begin{equation}\label{eq:uniform_projection}
    \delta_{\alpha'\gamma} = (1-r)\delta_{\lambda\gamma} + r\delta_{\lambda+s,\gamma}, \quad
    \delta_{\beta'\gamma} = (1-r)\delta_{\mu\gamma} + r\delta_{\mu-s,\gamma},
\end{equation}
where \(\bzeta_{\lambda+s} + \bzeta_{\mu-s} = \bzeta_{\lambda} + \bzeta_{\mu}\) and
\begin{equation}\label{eq:stencil_weights2}
    r = \frac{E_0-E_1}{E_2-E_1}, \quad
    E_0 = \bzeta_{\alpha'}^2 + \bzeta_{\beta'}^2, \quad
    E_1 = \bzeta_{\lambda}^2 + \bzeta_{\mu}^2, \quad
    E_2 = \bzeta_{\lambda+s}^2 + \bzeta_{\mu-s}^2.
\end{equation}
Subindex \(\nu\) is omitted for brevity.
For this scheme, the following relations hold:
\begin{equation}\label{eq:weights_ranges2}
    0 \leq r < 1, \quad h \leq R_{\mathcal{S}} \leq \sqrt3h,
\end{equation}
where \(h^3 = w_\gamma = V_\Gamma/|\Gamma|\).

Let \(\boldsymbol{\eta} = \bzeta_{\alpha'} - \bzeta_{\lambda}\),
and \(\bh_+\), \(\bh_-\) be the minimal diagonal displacements from \(\bzeta_{\lambda}\)
so that \(\bh_+\) is directed in the same octant as \(\boldsymbol{\eta}\)
and \(\bh_-\) lies in the opposite one.
Then, the \emph{compact 5-point scheme} is constructed on the nodes
\begin{equation}\label{eq:stencil_nodes5}
    \bzeta_{\lambda+s_0} = \bzeta_{\lambda}, \quad
    \bzeta_{\lambda+s_i} = \bzeta_{\lambda} + (\bh_+\cdot \be_i)\be_i, \quad
    \bzeta_{\lambda+s_4} = \bzeta_{\lambda} + \bh_-,
\end{equation}
where \(\be_i\) is the basis of the rectangular velocity grid. The projection weights are
\begin{equation}\label{eq:stencil_weights5}
    r_{\lambda,0} = 1 - \sum_{j=1}^4 r_{\lambda,j}, \quad
    r_{\lambda,i} = \frac{\eta_i - r_{\lambda,4}h_{-i}}{h_{+i}}, \quad
    r_{\lambda,4} = \frac{\boldsymbol{\eta}\cdot(\boldsymbol{\eta} - \bh_+)}
        {\bh_-\cdot(\bh_- - \bh_+)}.
\end{equation}
For the uniform grid, the following relations hold:
\begin{equation}\label{eq:weights_ranges5}
    0 < r_{\lambda,0} \leq 1, \quad
    -\frac1{12} \leq r_{\lambda,i} < \frac{11}{24}, \quad
    -\frac18 \leq r_{\lambda,4} \leq 0, \quad
    R_\mathcal{S} = \sqrt6h.
\end{equation}

The \emph{symmetric 7-point scheme} is constructed on the nodes
\begin{equation}\label{eq:stencil_nodes7}
    \bzeta_{\lambda+s_0} = \bzeta_{\lambda}, \quad
    \bzeta_{\lambda+s_{\pm i}} = \bzeta_{\lambda} + (\bh_\pm\cdot \be_i)\be_i.
\end{equation}
The projection weights are
\begin{equation}\label{eq:stencil_weights7}
    r_{\lambda,0} = 1 - \sum_{j=1}^3 r_{\lambda,j} + r_{\lambda,-j}, \quad
    r_{\lambda,\pm i} = \pm\frac{\eta_i(\eta_i - h_{\mp i})}{h_{\pm i}(h_{+i}-h_{-i})}.
\end{equation}
There is no summation over repeated indices in~\eqref{eq:stencil_weights7}.
For the uniform grid, the following relations hold:
\begin{equation}\label{eq:weights_ranges7}
    \frac14 \leq r_{\lambda,0} \leq 1, \quad
    0 \leq r_{\lambda,\pm i} \leq \frac38, \quad
    -\frac18 \leq r_{\lambda,\mp i} \leq 0, \quad
    R_\mathcal{S} = 2h.
\end{equation}

Both 5-point and 7-point schemes have \(\bar{r}_{\max}=1/8\).
In order to reduce it, more projection nodes are required~\cite{Dodulad2012}.

\section{Numerical examples}

In the present work, solver \verb+snitSimpleFoam+~\cite{Rogozin2014}, developed
within the open-source computational platform OpenFOAM\textregistered{}, is employed for the KGF equations.
The solver is extended to deal with the second-order boundary conditions.
Grids are adjusted in such a way as to ensure the numerical error is less than \(10^{-4}\).
The iterative process is stopped when the residual becomes less than \(10^{-6}\).

The described algorithm is implemented within the problem-solving environment
for numerical analysis of rarefied gas flows~\cite{Kloss2010, Kloss2012}.
The physical grid is selected under the same criteria as for the numerical solution of the KGF equations,
but needs additional exponential refinement near the diffuse-reflection boundaries
to ensure the numerical error is less than \(10^{-4}\) in the Knudsen layer.
Both uniform and nonuniform velocity grids are used, while simulation on a detailed grid
starts from a solution on a coarse grid to accelerate the steady-state convergence.
For nonuniform grids, the symmetric 7-point scheme is employed,
since it provides smaller \(\epsilon_J\) for a smooth distribution function.
The number of cubature points is adjusted so that \(\epsilon_J < 10^{-5}\).

\subsection{Flow between the parallel plates with the temperature distributed sinusoidally}

\begin{wrapfigure}{r}{7.4cm}
    \vspace{-20pt}
    \centering
    \includegraphics{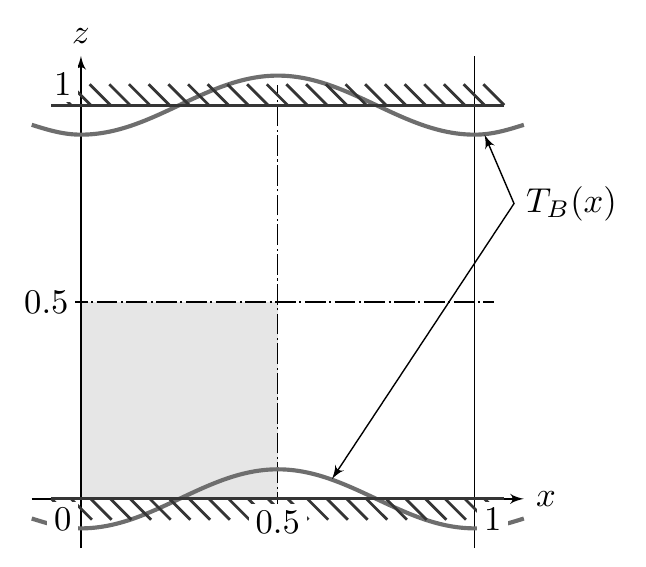}
    \vspace{-10pt}
    \caption{Geometry of the problem}\label{fig:sone_bobylev}
\end{wrapfigure}

Consider a plane periodic geometry, as shown in Fig.~\ref{fig:sone_bobylev}.
Gas is placed between the two parallel plates at rest (\(v_{Bi} = 0\)) separated by unit distance.
Their temperature is distributed sinusoidally:
\begin{equation}
    T_B = 1 - \frac{\cos(2\pi x)}{2}.
\end{equation}
The complete diffuse-reflection boundary condition is used at the plates.
The gas density is normalized to unity, i.e.,
\begin{equation}\label{eq:total_mass}
    \int_0^1\int_0^1\rho\dd{x}\dd{y} = 1.
\end{equation}
Due to the symmetry of the problem, the computational domain is the square with the side \(1/2\).
It is grayed out in Fig.~\ref{fig:sone_bobylev}.

This problem was studied in~\cite{Sone1996} based both on the KGF equations and on the kinetic approach;
however, only model Krook--Welander equation was exploited due to the high complexity
of the numerical approximation of the Boltzmann equation.
For gas mixtures, some numerical results are presented in~\cite{Wu2015}.
In the present paper, a direct solution of the Boltzmann equation is obtained for a hard-sphere gas.
The asymptotic solution for small \(\Kn\) is examined as part of the problem.
In particular, a comparative analysis is performed for the high-order boundary conditions.

\subsubsection{Solution in the continuum limit}

\begin{figure}
    \centering
    \subfloat[the heat-conduction equation~\eqref{eq:heat_equation}]{
        \includegraphics[width=0.5\textwidth]{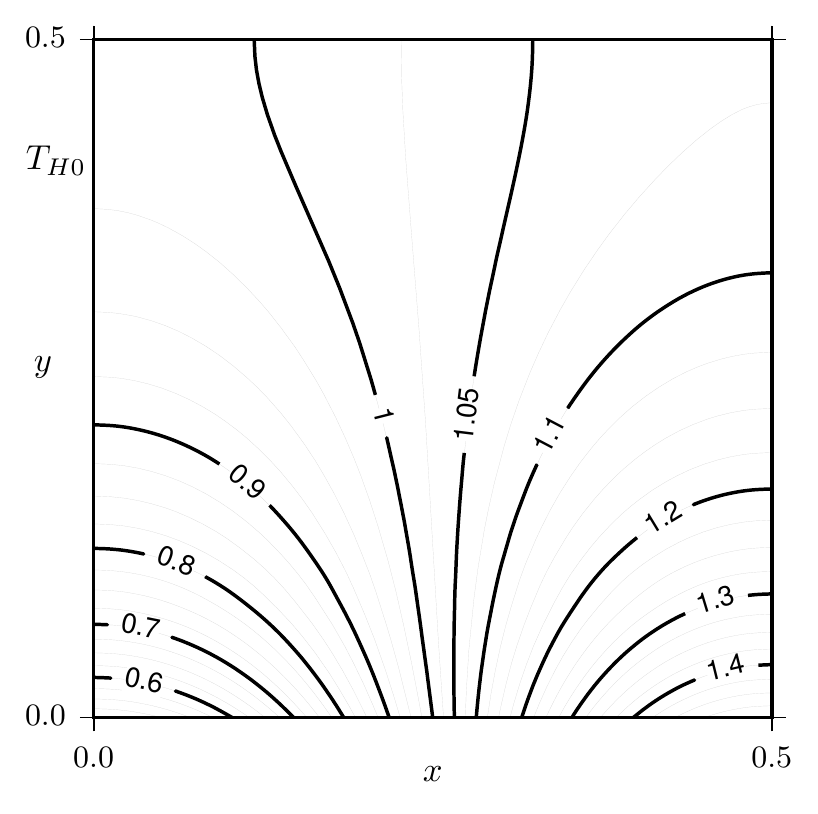}
        \label{fig:continuum:temp-heat}
    }
    \subfloat[the KGF equations]{
        \includegraphics[width=0.5\textwidth]{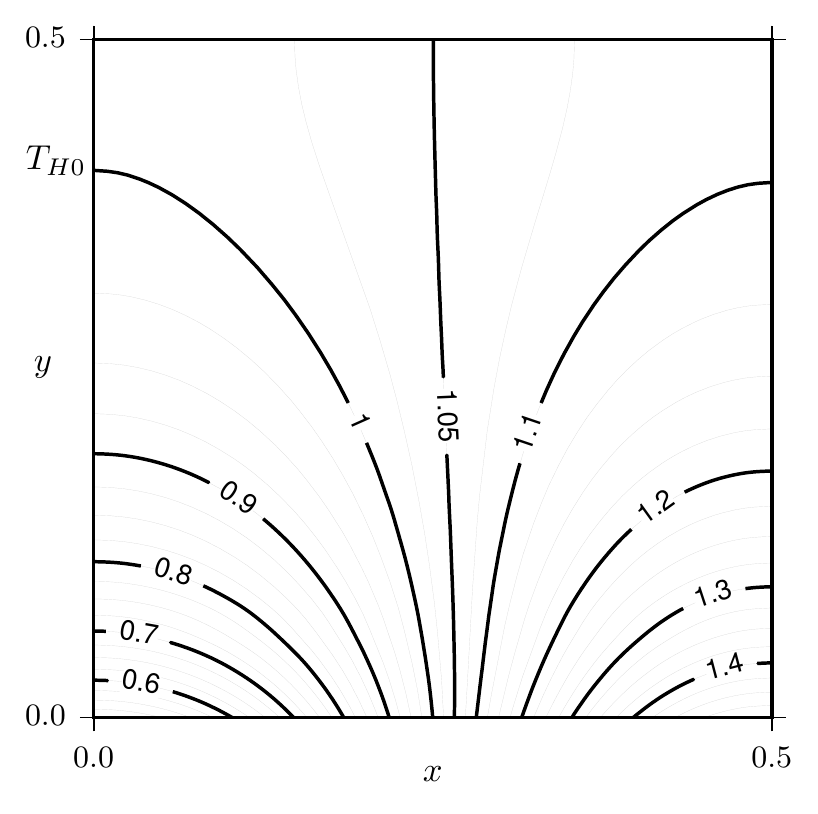}
        \label{fig:continuum:temp-snit}
    }
    \caption{Isothermal lines in the continuum limit}
    \label{fig:continuum:temp}
\end{figure}

\begin{figure}
    \centering
    \subfloat[the KGF equations without thermal creep (\(K_1=0\))]{
        \includegraphics[width=0.5\textwidth]{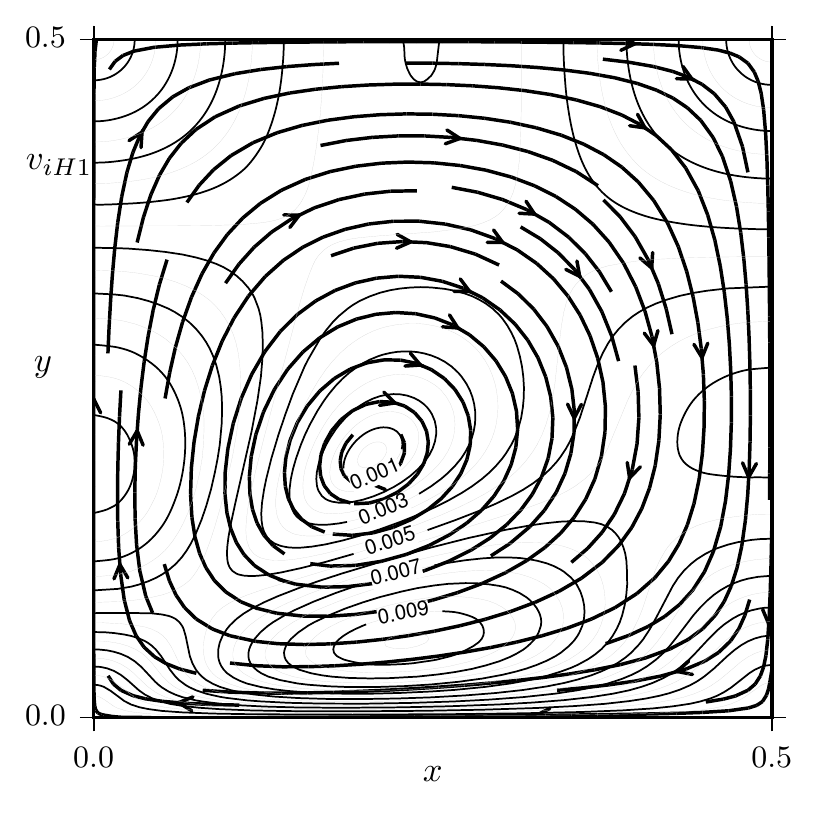}
        \label{fig:continuum:flow-nonslip}
    }
    \subfloat[the KGF equations]{
        \includegraphics[width=0.5\textwidth]{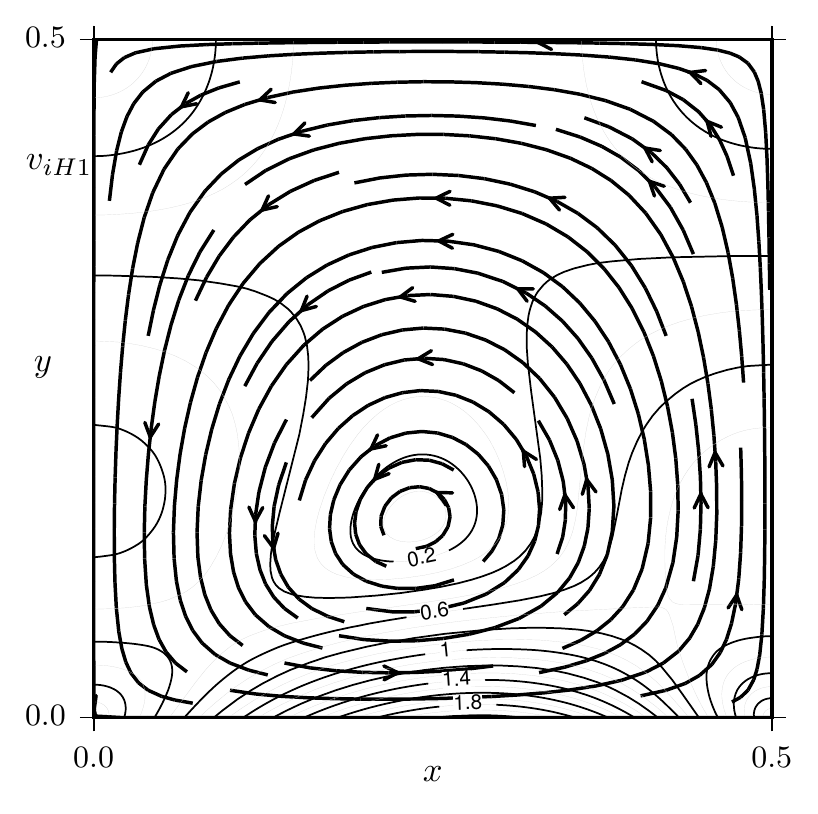}
        \label{fig:continuum:flow-snit}
    }
    \caption{The velocity field \(v_{i1}\) in a continuum limit.
        Contour lines correspond to the magnitude, and curves with arrows show the direction.}
    \label{fig:continuum:flow}
\end{figure}

For the numerical analysis of the problem, the rectangular grid is used in the physical space:
the region \(0<x<1/2\) is divided into 30 intervals of equal length,
and the region \(0<y<1/2\), into 40 intervals with refinement near \(y=0\).

Fig.~\ref{fig:continuum:temp} shows the steady-state temperature field
obtained as the solution of the heat-conduction and KGF equations.
In the continuum limit, the heat-conduction equation is derived from the KGF equations,
when \(\bar{\gamma}_7=0\) and \(K_1=0\).
The thermal-creep flow is much stronger than the nonlinear thermal-stress one.
This fact is demonstrated in Fig.~\ref{fig:continuum:flow},
where the KGF equations are solved both with and without the thermal-creep boundary condition.
Note also that direction of the gas flows are opposite in Fig.~\ref{fig:continuum:flow-nonslip} and~\ref{fig:continuum:flow-snit}.
The results obtained in the continuum limit coincides with the presented in~\cite{Sone1996}.
This fact can serve as a verification of the solver \verb+snitSimpleFoam+.

\subsubsection{Solution for arbitrary Knudsen numbers}

\setlength{\tabcolsep}{4pt}
\begin{table}
    \caption{The parameters of the velocity grids:
        \(\zeta_{\mathrm{cut}}\) is the radius of the sphere that encloses all nodes,
        \(N_i\) is the maximum number of nodes along the \(x_i\) axis,
        \(|\mathcal{V}|\) is the total number of nodes,
        \(\min(\Delta\zeta_i)\) is the minimum distance between nodes along the \(x_i\) axis,
        \(\delta T_M/T\) is the relative temperature error of the Maxwellian,
        \(\delta T_{MM}/T\) is the relative temperature error of the sum of two half-Maxwellians,
        separated by the plane \(\zeta_y=0\)}
    \label{table:velocity_grids}
    \centering
    \begin{tabular}{ccccccccc}
        Grid & \(\zeta_{\mathrm{cut}}\) & \(N_{x,z}/2\) & \(N_y/2\) & \(|\mathcal{V}|\)
            & \(\min(\Delta\zeta_{x,z})\) & \(\min(\Delta\zeta_y)\) & \(\delta T_M/T\) & \(\delta T_{MM}/T\) \\\hline
        M1 & 4.25 & 8  & 8  & 2176  & 0.53 & 0.53  & \([-20,0.3]\cdot10^{-5}\)   & \(-[4.0,10]\cdot10^{-5}\) \\
        M2 & 5.3  & 11 & 26 & 20248 & 0.4  & 0.005 & \([2.3,4.1]\cdot10^{-5}\)   & \([2.7,4.0]\cdot10^{-5}\) \\
        M3 & 4.5  & 12 & 15 & 15568 & 0.05 & 0.005 & \([4.0,5.1]\cdot10^{-3}\)   & \([4.4,5.5]\cdot10^{-3}\) \\
        M4 & 8.0  & 16 & 16 & 28640 & 0.1  & 0.1   & \([1.65,1.79]\cdot10^{-3}\) & \([3.1,3.4]\cdot10^{-3}\) \\
    \end{tabular}
\end{table}

\begin{figure}
    \centering
    \subfloat[The KGF equations with the second-order boundary conditions]{
        \includegraphics[width=0.5\textwidth]{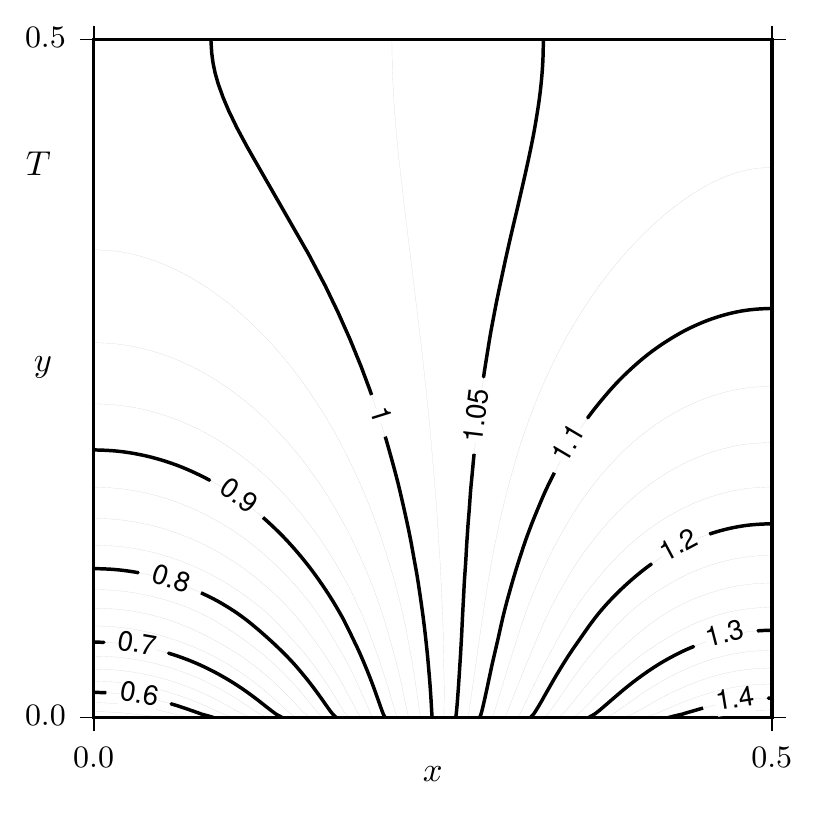}
        \label{fig:kn0.01:temp-snit}
    }
    \subfloat[the Boltzmann equation]{
        \includegraphics[width=0.5\textwidth]{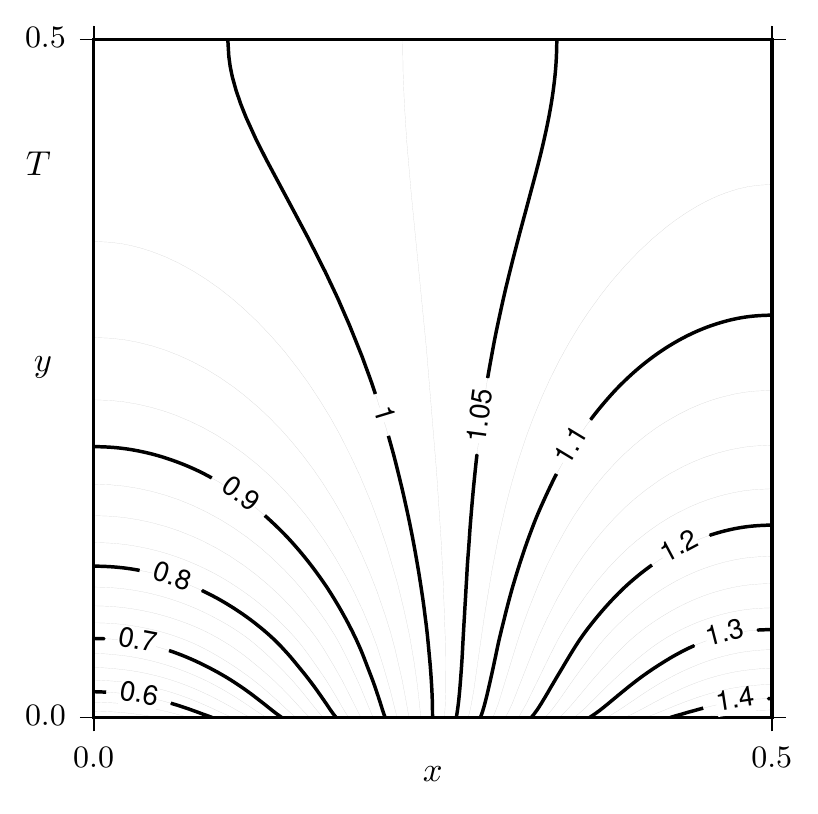}
        \label{fig:kn0.01:temp-exact}
    }
    \caption{Isothermal lines for \(\Kn=0.01\)}
    \label{fig:kn0.01:temp}
\end{figure}

\begin{figure}
    \centering
    \subfloat[the KGF equations with the second-order boundary conditions]{
        \includegraphics[width=0.5\textwidth]{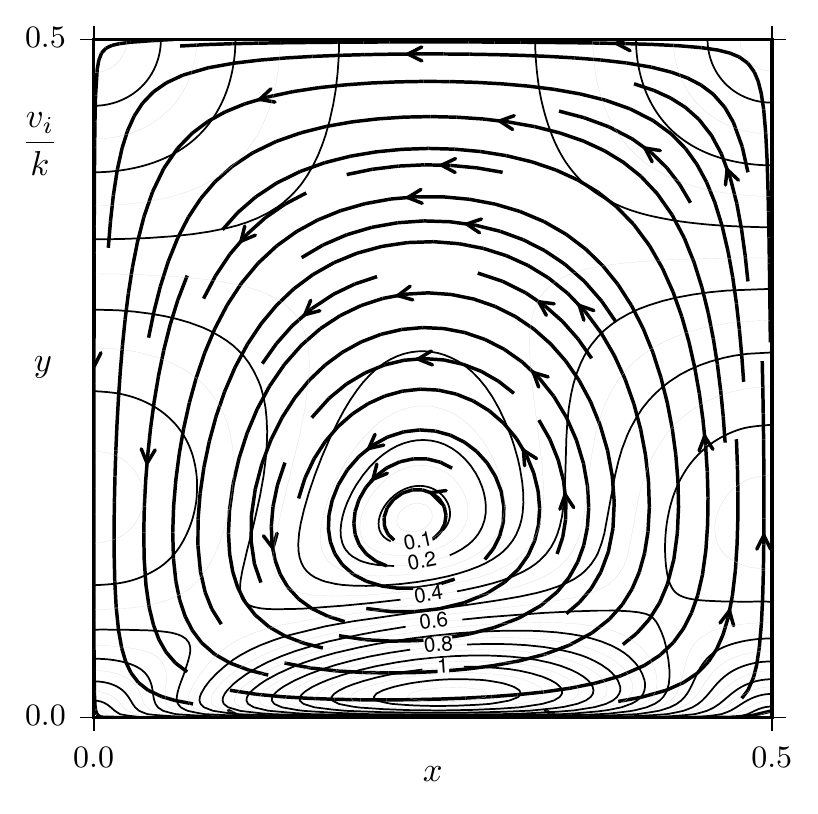}
        \label{fig:kn0.01:flow-snit}
    }
    \subfloat[the Boltzmann equation]{
        \includegraphics[width=0.5\textwidth]{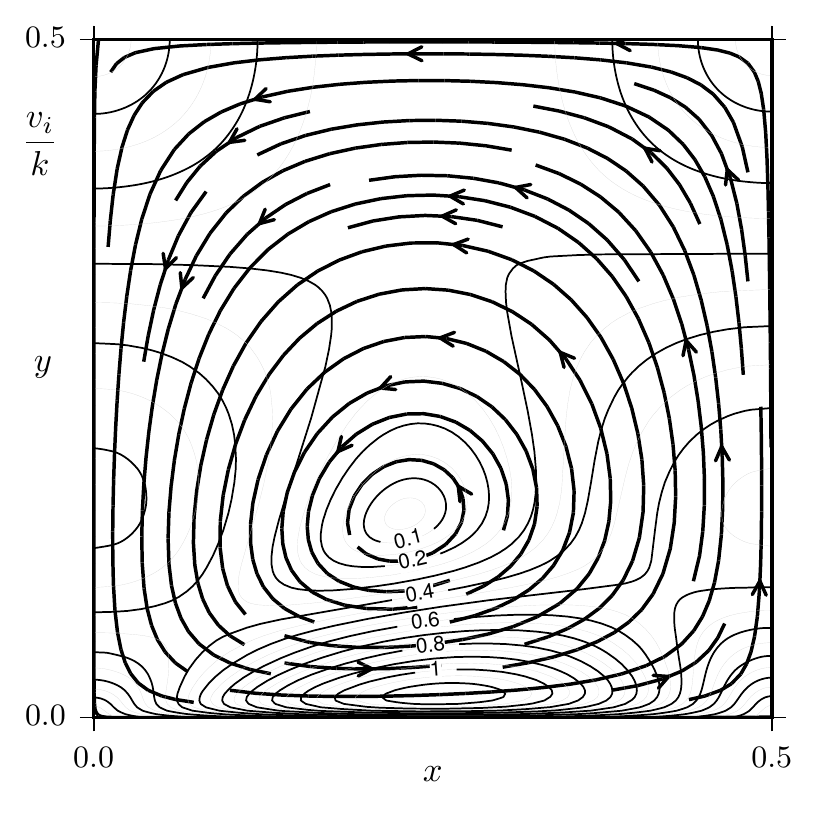}
        \label{fig:kn0.01:flow-exact}
    }
    \caption{The velocity field for \(\Kn=0.01\).
        Contour lines correspond to the magnitude, and curves with arrows show the direction.}
    \label{fig:kn0.01:flow}
\end{figure}

\begin{figure}
    \centering
    \subfloat[the temperature field]{
        \includegraphics[width=0.5\textwidth]{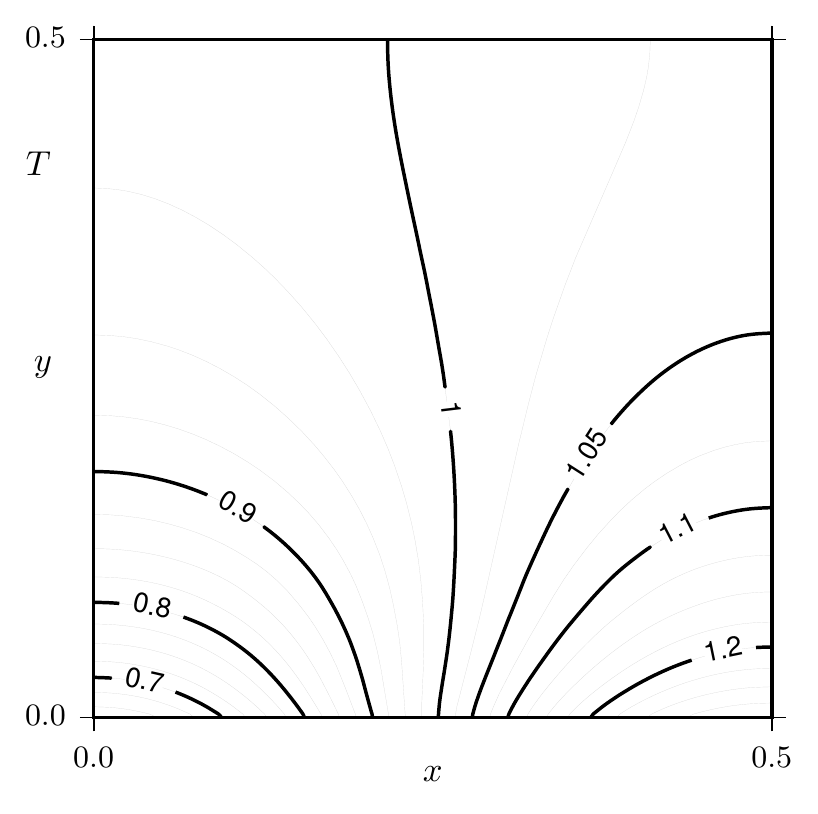}
        \label{fig:kn0.1:temp}
    }
    \subfloat[the velocity field]{
        \includegraphics[width=0.5\textwidth]{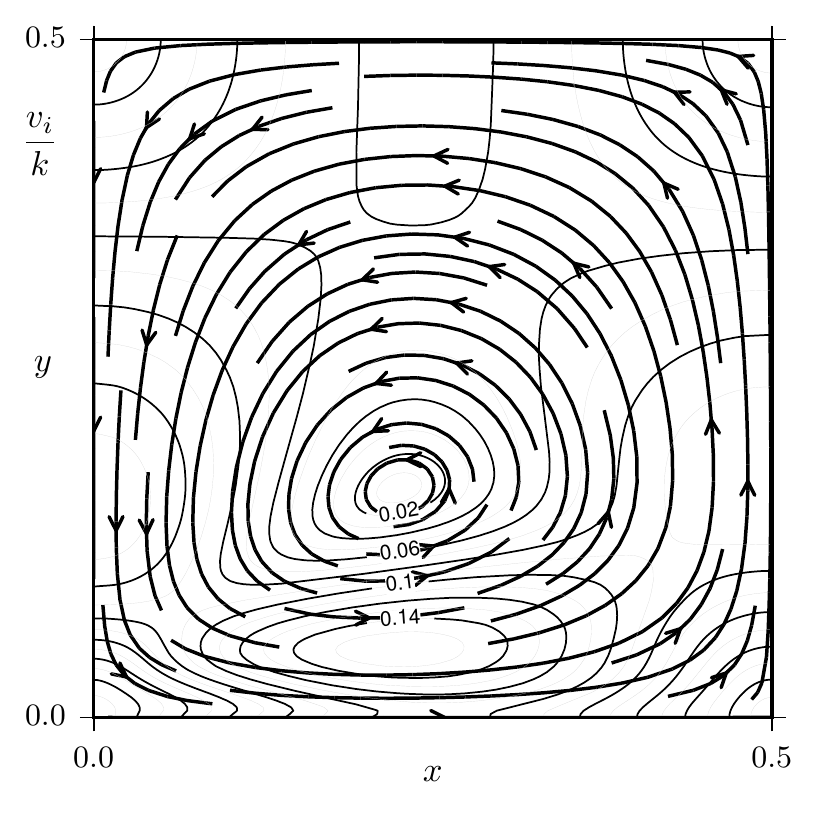}
        \label{fig:kn0.1:flow}
    }
    \caption{The solution of the Boltzmann equation for \(\Kn=0.1\).
        Contour lines correspond to the magnitude, and curves with arrows show the direction.}
    \label{fig:kn0.1}
\end{figure}

To consider the problem in arbitrary range of Knudsen numbers,
it is necessary to refer to the numerical solution of the Boltzmann equation.
The physical grid is the same as in the solution of KGF equations;
however, it is additionally refined in the Knudsen layer (close to \(y=0\))
so that the width of the boundary cell is equal to \(0.02\) mean free paths.

The numerical accuracy in the velocity space is controlled on several rectangular grids (M1, M2, M3),
whose parameters are presented in Table~\ref{table:velocity_grids}.
The problem is initially solved on the coarse uniform grid M1,
and then the result is refined on the nonuniform grids.
M2 nodes are arranged as the roots of the Hermite polynomial along the \(x\) and \(z\) axes,
exponentially along the \(y\) axis. Such a strong refinement provides a careful approximation
of the large variations of the distribution function in the Knudsen layer.
The distance between M3 nodes grows quadratically along each axis.
Unlike M2, M3 approximates cold distributions (a temperature close to \(T=0.5\)) more accurately.
Number of cubature points varies considerably:
about \(5\times 10^3\) for M1 and about \(5\times 10^4\) for nonuniform grids.
For very small \(\Kn\), the macroscopic fields are computed from time extrapolation instead of time averaging,
since the achievement of the completely steady state (especially near \(y=1/2\))
requires too many iterations of the explicit scheme.

For nonuniform grids, it is important to check the computational accuracy of the macroscopic variables
(some cubatures of the form~\eqref{eq:bzeta_cubature}).
For example, the ranges of the relative temperature error for the Maxwellian (\(\delta T_M\))
and the sum of the two half-Maxwellians (\(\delta T_{MM}\)) are shown in Table~\ref{table:velocity_grids}.
They are computed for the typical temperatures and velocities.
Moments of a distribution function are calculated accurately on a coarse grid,
if its nodes are arranged uniformly or as the roots of the Hermite polynomial.
Otherwise, the velocity grid should be refined sufficiently as it is done along the \(y\) axis for M2.
For M3, the temperature cubature~\eqref{eq:bzeta_cubature} is always about \(0.004\) more than the true value
and therefore can be adjusted in accordance with this value.
For M1, the cubature error takes negative values for temperatures close to \(T=1.5\),
because \(\zeta_{\mathrm{cut}}\) is not sufficiently large.

Fig.~\ref{fig:kn0.01:temp} and~\ref{fig:kn0.01:flow} show the temperature and velocity fields for \(\Kn=0.01\).
The numerical solutions of the Boltzmann and KGF equations can be compared for small \(\Kn\).
It is clearly seen that the temperature field obtained with the high-order boundary conditions (Fig.~\ref{fig:kn0.01:temp-snit})
significantly better approximates the exact solution (Fig.~\ref{fig:kn0.01:temp-exact})
compared with the solution obtained with the leading-order ones (Fig.~\ref{fig:continuum:temp-snit}).

Fig.~\ref{fig:kn0.1} represents the solution for \(\Kn=0.1\).
With the increase in \(\Kn\), the thermal-creep flow and temperature jump near the boundary \(y=0\) amplify.
The point of maximum gas velocity moves away from the plate.

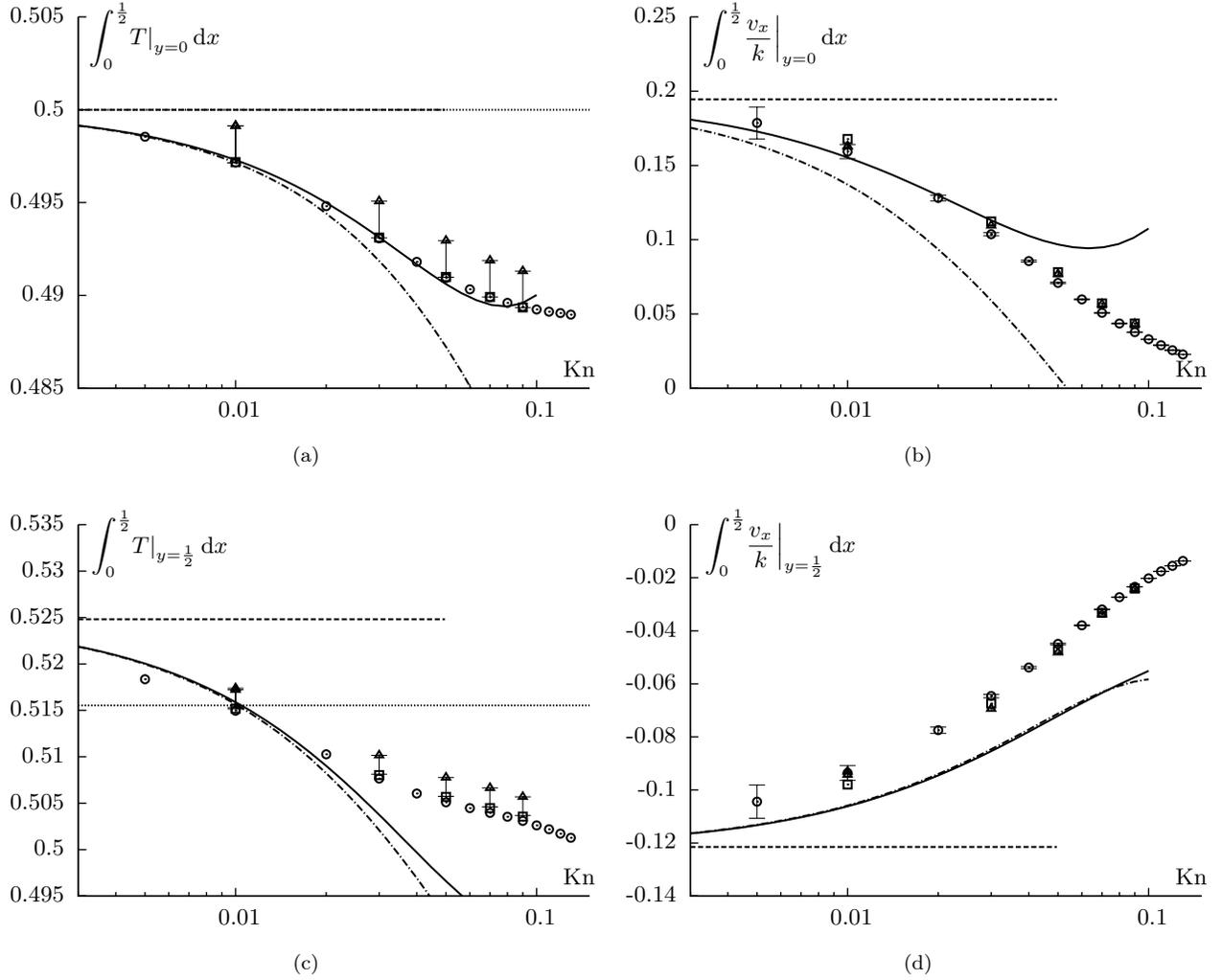
\begin{figure}
    \scalebox{0}{
        \begin{tikzpicture}
            \begin{axis}[hide axis]
                \addplot[line width=0.8, dotted](0,0);                      \label{leg:heat}
                \addplot[line width=0.8, dashed](0,0);                      \label{leg:asym0}
                \addplot[line width=0.8, dashdotted](0,0);                  \label{leg:asym1}
                \addplot[line width=0.8, solid](0,0);                       \label{leg:asym2}
                \addplot[line width=0.8, only marks, mark=o](0,0);          \label{leg:data1}
                \addplot[line width=0.8, only marks, mark=square](0,0);     \label{leg:data2}
                \addplot[line width=0.8, only marks, mark=triangle](0,0);   \label{leg:data3}
            \end{axis}
        \end{tikzpicture}
    }%
    \centering
    \subfloat[]{
        \includegraphics[width=0.5\textwidth]{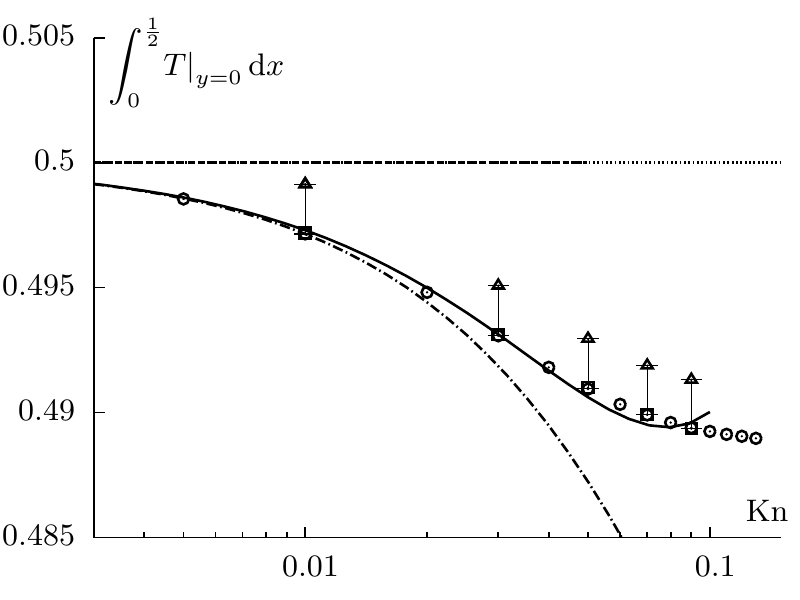}
        \label{fig:comparison:bottomT}
    }
    \subfloat[]{
        \includegraphics[width=0.5\textwidth]{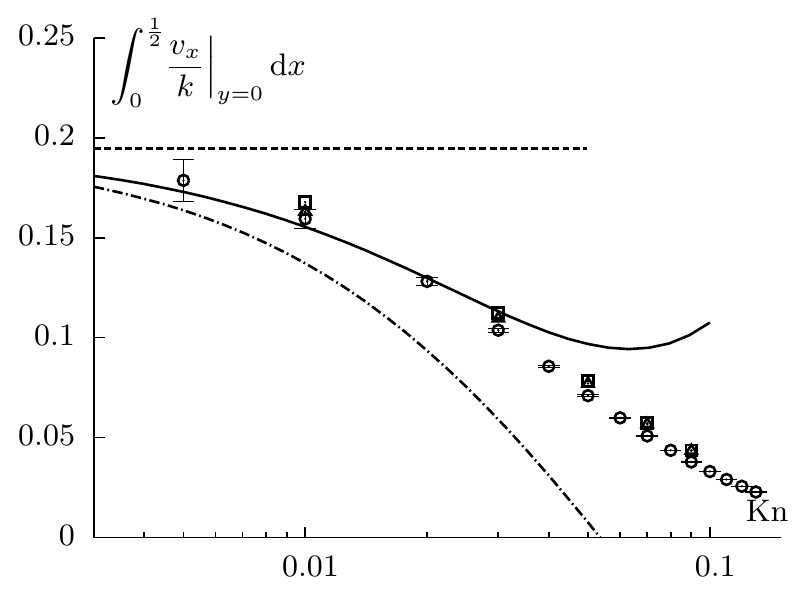}
        \label{fig:comparison:bottomU}
    }\\
    \subfloat[]{
        \includegraphics[width=0.5\textwidth]{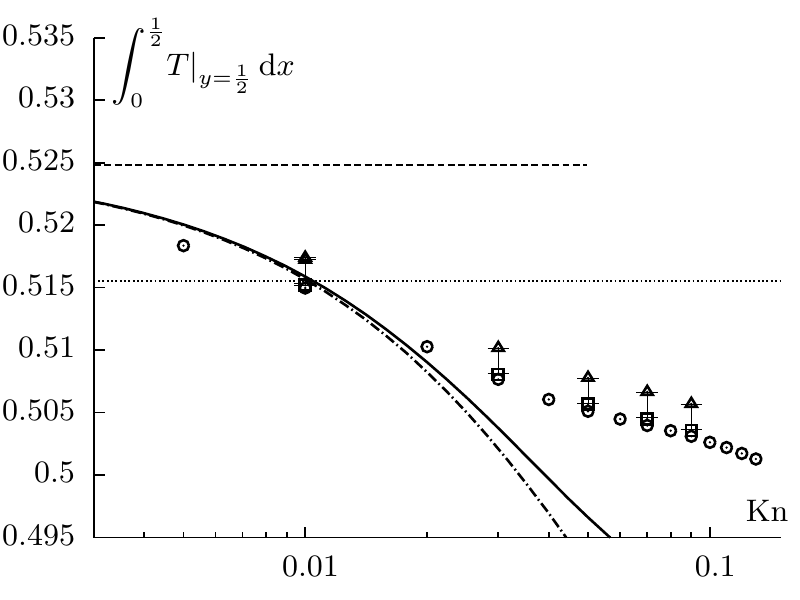}
        \label{fig:comparison:topT}
    }
    \subfloat[]{
        \includegraphics[width=0.5\textwidth]{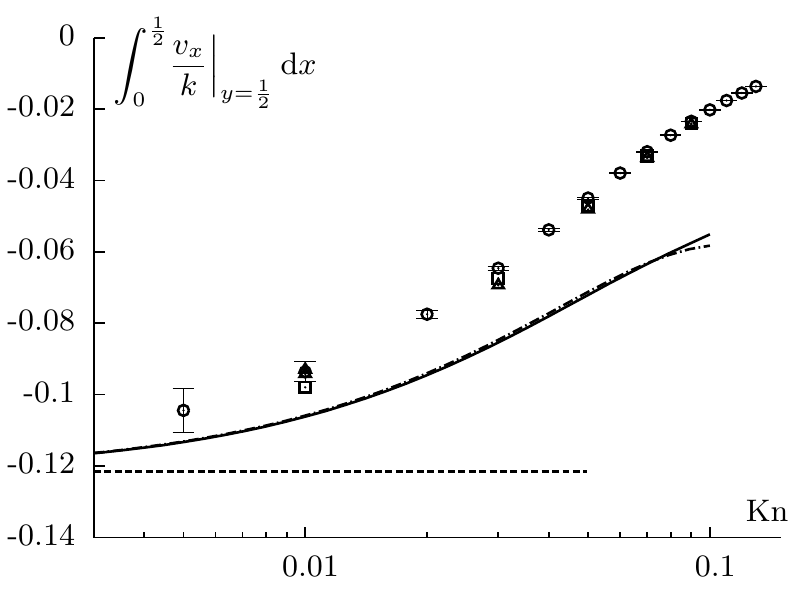}
        \label{fig:comparison:topU}
    }
    \caption{
        Some boundary integrals, obtained by different methods:
        the heat-conduction equation~\ref{leg:heat},
        the KGF equations with the leading-order (only thermal creep)~\ref{leg:asym0},
        first-order~\ref{leg:asym1}, and second-order~\ref{leg:asym2} boundary conditions,
        the Boltzmann equation on grids M1~\ref{leg:data1}, M2~\ref{leg:data2}, and M3~\ref{leg:data3}.
        Error bars for M3 correspond to the temperature correction in accordance with
        the cubature error for a Maxwellian.
        Error bars for M1 correspond to the relative error \(3\times 10^{-4}\).
    }
    \label{fig:comparison}
\end{figure}

To demonstrate the convergence of the Boltzmann solution to the KGF one in a continuum limit,
some boundary integrals versus the Knudsen number are presented in Fig.~\ref{fig:comparison}.
It is seen from Fig.~\ref{fig:comparison:bottomT} that all boundary conditions for the KGF equations,
together with the corresponding Knudsen-layer corrections,
approximate the numerical solution of the Boltzmann equation with the stated accuracy.
In particular, the condition~\eqref{eq:boundary_T0} yields the error \(\OO{k}\),
the first-order temperature jump leads to \(\OO{k^2}\),
and the second-order jump improves the convergence up to \(\OO{k^3}\).
Note that the solution obtained on the coarse uniform grid M1 is almost identical to the M2 solution
and to the adjusted M3 one.

In Fig.~\ref{fig:comparison:bottomU}, the same convergence rates are observed for \(v_i\).
Since \(v_i/k\) is depicted, the error of the numerical solution of the Boltzmann solution increased for small \(k\).
The M2 and M3 solutions are almost identical, but differ by a constant value (about \(0.008\)) from the M1 solution.
Thus, discontinuities, as well as their decay in the Knudsen layer, of the distribution function
on the diffuse-reflection boundary make a negligible contribution to the overall solution.
For slow flows, it is explained by the fact that these discontinuities are \(\OO{k}\).
This observation allows us to consider the distribution function is smooth enough without significant loss in accuracy;
therefore, it is possible to avoid adaptation of the velocity grid to the geometry of the problem.

The fluid-dynamic and kinetic solutions differ, as expected, on the magnitude \(\OO{k}\) away from the diffuse-reflection boundary.
It is demonstrated in Fig.~\ref{fig:comparison:topT} and~\ref{fig:comparison:topU}.
It is clearly seen from Fig.~\ref{fig:comparison:topT} that the solution of the Boltzmann equation
converges to the solution of the KGF equations rather than the heat-conduction equation.
The corrected M3 solution is almost identical with M2 and slightly exceeds M1 (about 0.001).
This difference seems to be explained by a rough approximation of the M1 grid.
It is also seen from Fig.~\ref{fig:comparison:topT} and~\ref{fig:comparison:topU} that the second-order boundary conditions
have little effect on the solution of KGF equations;
however, the accuracy of the asymptotic solution is significantly improved if only the temperature and velocity jumps
are included in the boundary conditions.

\subsection{Flow between two uniformly heated elliptic cylinders}

\begin{figure}
    \centering
    \subfloat[the KGF equations with the leading-order boundary conditions]{
        \includegraphics{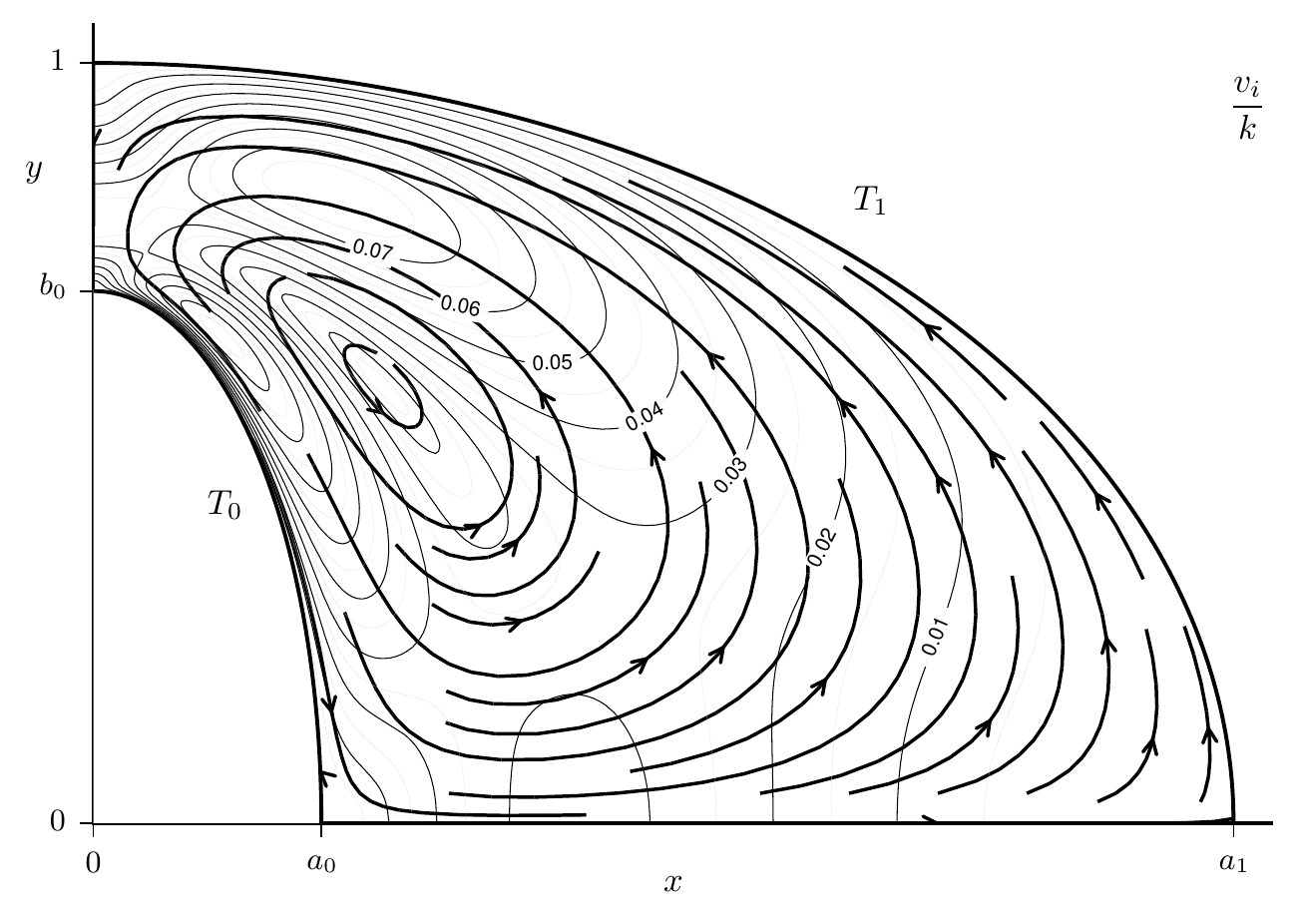}
        \label{fig:elliptic:flow-kgf}
    }\\
    \subfloat[the KGF equations with the second-order boundary conditions]{
        \includegraphics{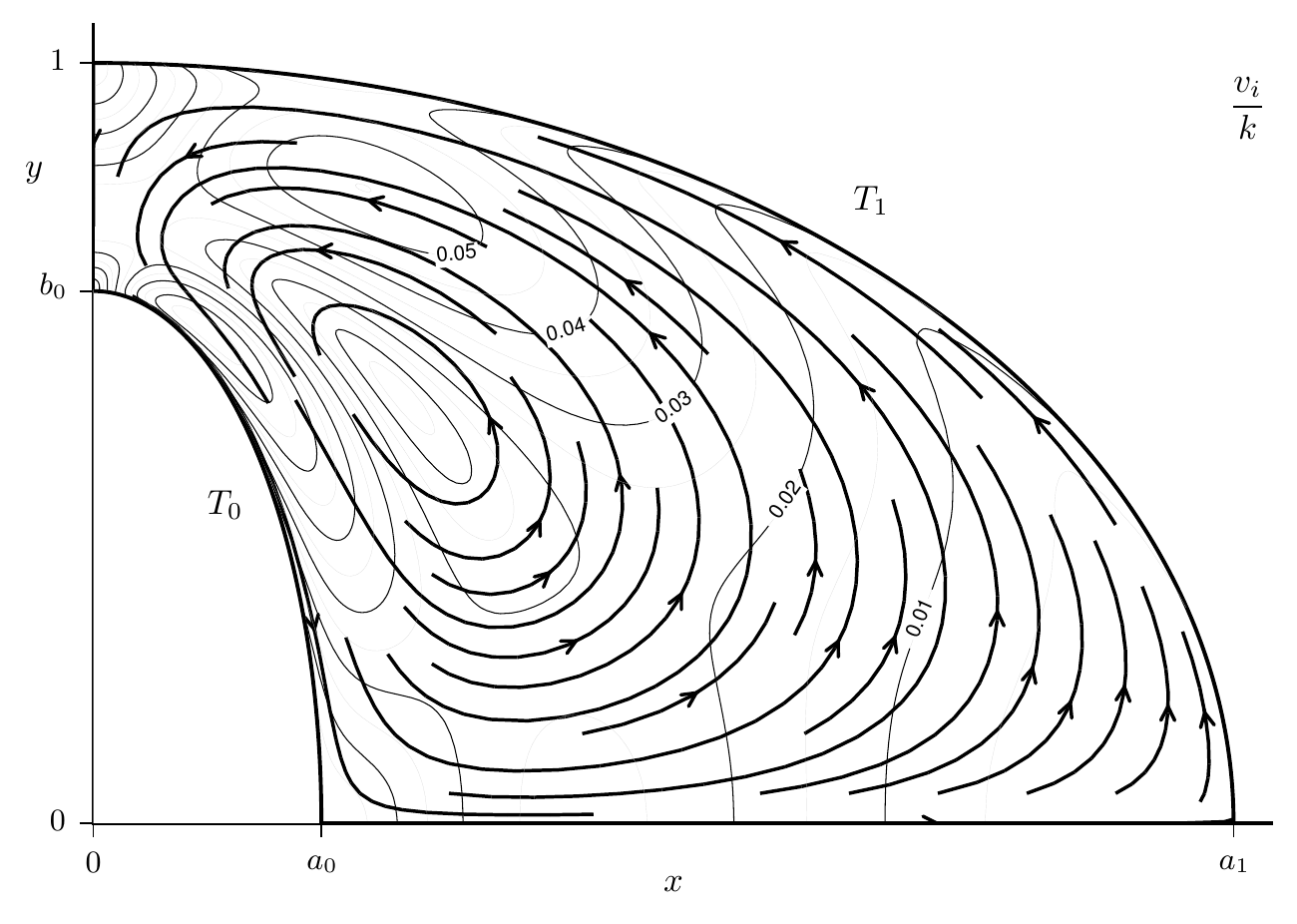}
        \label{fig:elliptic:flow-asym}
    }
    \caption{The velocity field for \(\Kn=0.02\).
        Contour lines correspond to the magnitude, and curves with arrows show the direction.}
\end{figure}

\begin{figure}
    \ContinuedFloat 
    \subfloat[the Boltzmann equation]{
        \includegraphics{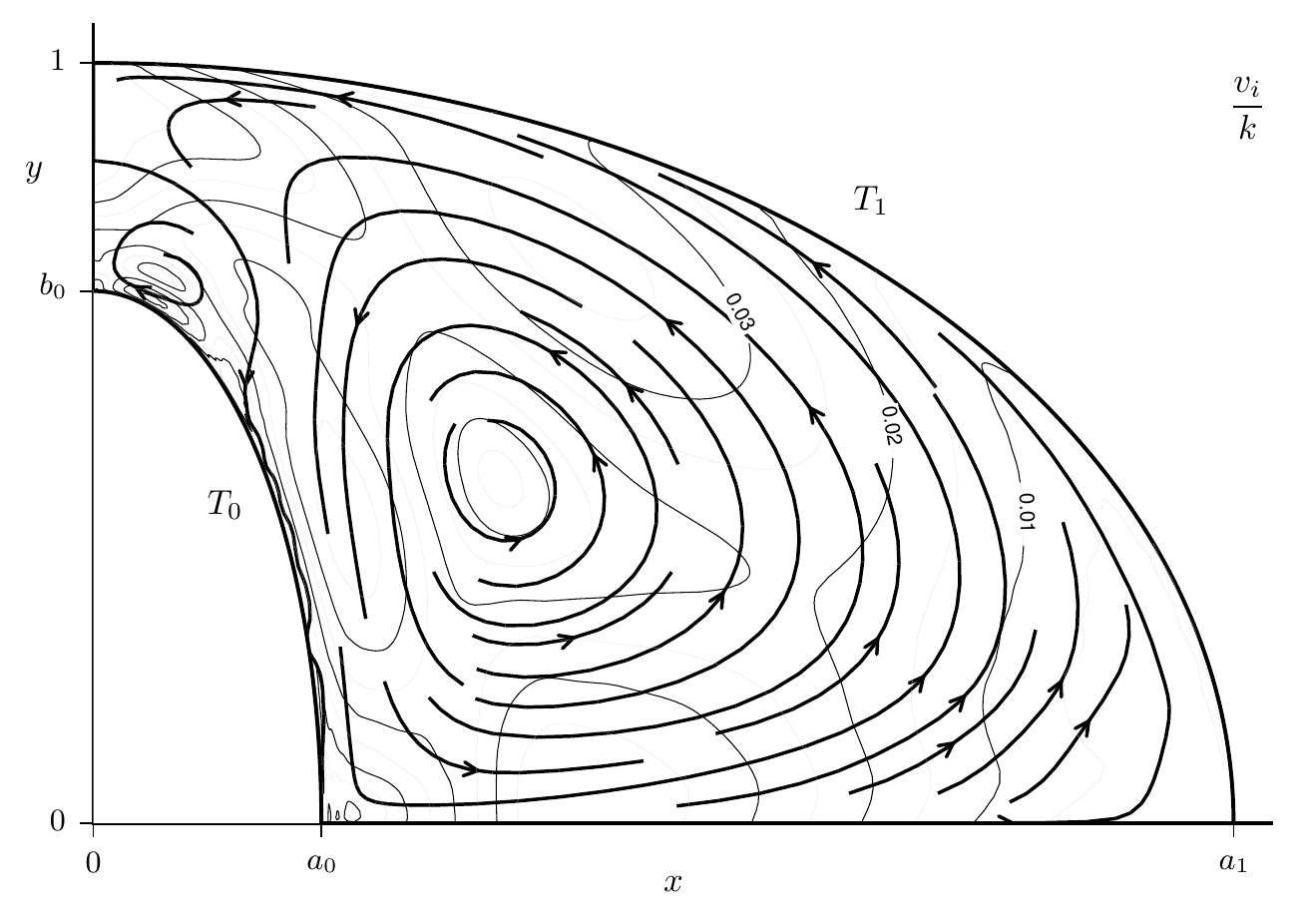}
        \label{fig:elliptic:flow-kes}
    }
    \caption{(cont.) The velocity field for \(\Kn=0.02\).
        Contour lines correspond to the magnitude, and curves with arrows show the direction.}
    \label{fig:elliptic:flow}
\end{figure}

\begin{figure}
    \setcounter{subfigure}{0}
    \scalebox{0}{
        \begin{tikzpicture}
            \begin{axis}[hide axis]
                \addplot[line width=0.8, solid](0,0);       \label{leg:kes}
                \addplot[line width=0.8, dotted](0,0);      \label{leg:kgf}
                \addplot[line width=0.8, dashed](0,0);      \label{leg:first}
                \addplot[line width=0.8, dashdotted](0,0);  \label{leg:second}
            \end{axis}
        \end{tikzpicture}
    }%
    \centering
    \subfloat[the outer ellipse]{
        \includegraphics[width=0.5\textwidth]{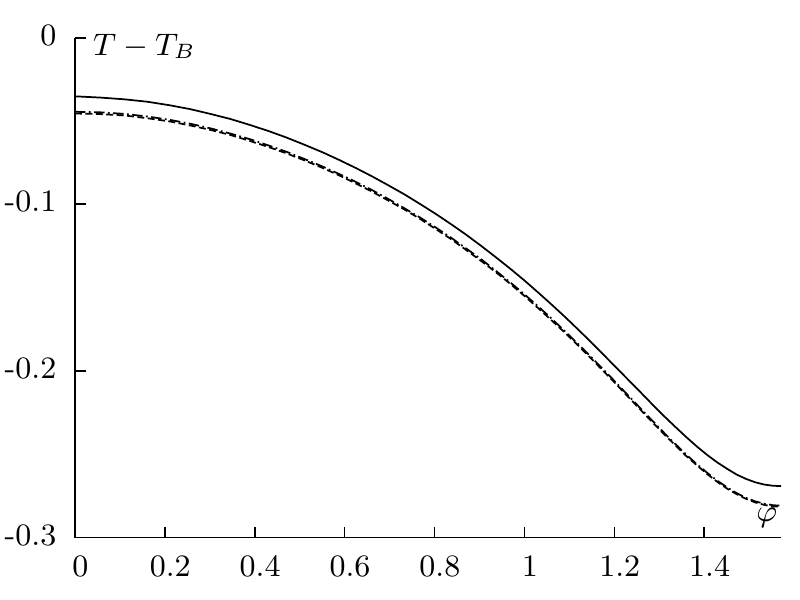}
        \label{fig:profile-temp:outer}
    }
    \subfloat[the inner ellipse]{
        \includegraphics[width=0.5\textwidth]{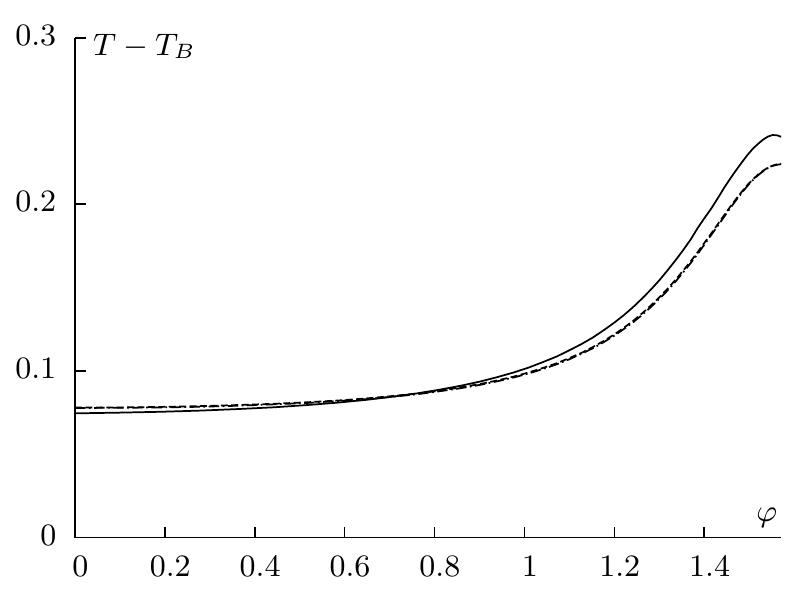}
        \label{fig:profile-temp:inner}
    }\\
    \subfloat[the semi-major axis of the outer cylinder]{
        \includegraphics[width=0.5\textwidth]{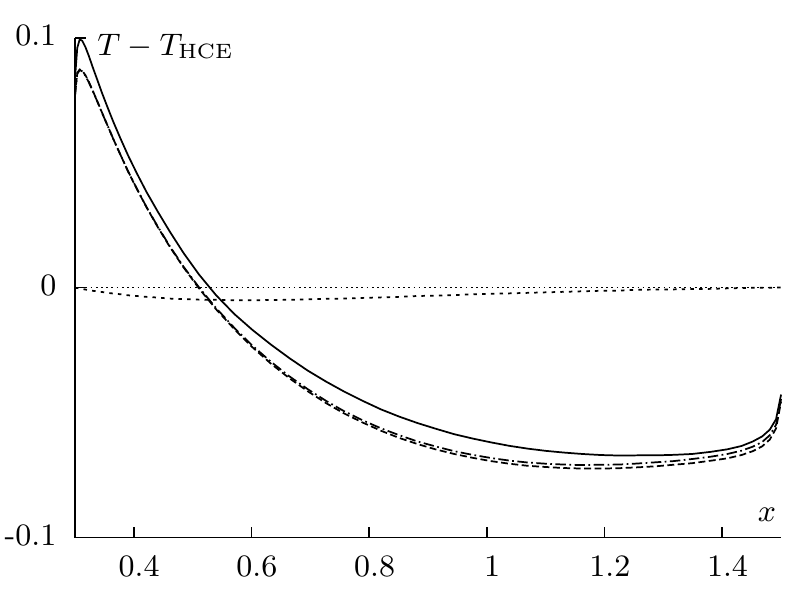}
        \label{fig:profile-temp:bottom}
    }
    \subfloat[the semi-minor axis of the outer cylinder]{
        \includegraphics[width=0.5\textwidth]{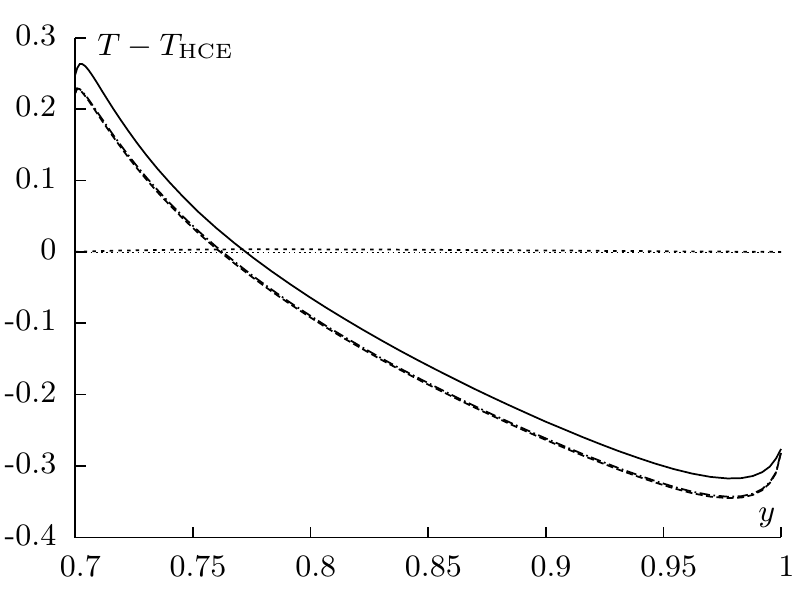}
        \label{fig:profile-temp:left}
    }
    \caption{
        The profile of the boundary temperature. The angle \(\varphi\) corresponds
        to the polar coordinates \(x=r\cos\varphi\), \(y=r\sin\varphi\).
        \(T_{\mathrm{HCE}}\) is the temperature field, described by the heat-conduction equation.
        The following solutions are presented: the Boltzmann equation~\ref{leg:kes} (divided by \(1.0017\)),
        the KGF equations with the leading-order~\ref{leg:kgf}, first-order~\ref{leg:first},
        and second-order~\ref{leg:second} boundary conditions.
    }
    \label{fig:profile-temp}
\end{figure}

\begin{figure}
    \centering
    \subfloat[the outer ellipse]{
        \includegraphics[width=0.5\textwidth]{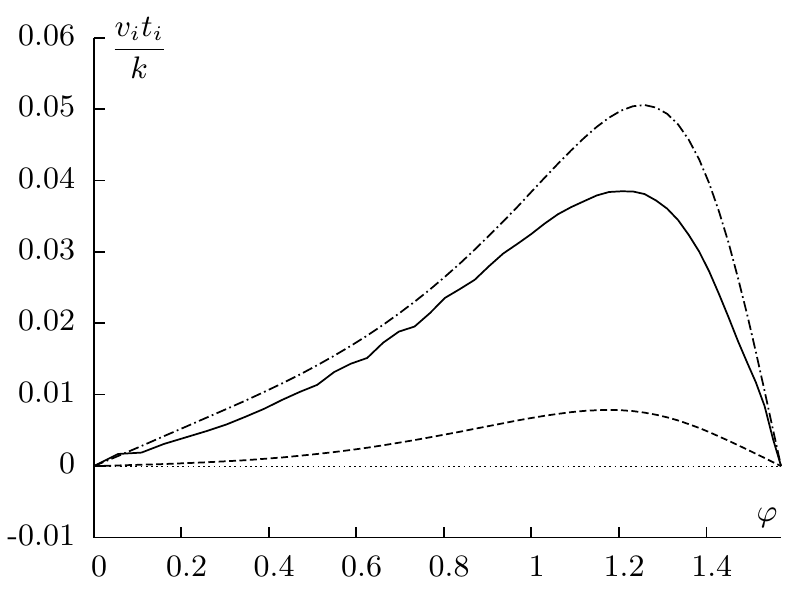}
        \label{fig:profile-vel:outer}
    }
    \subfloat[the inner ellipse]{
        \includegraphics[width=0.5\textwidth]{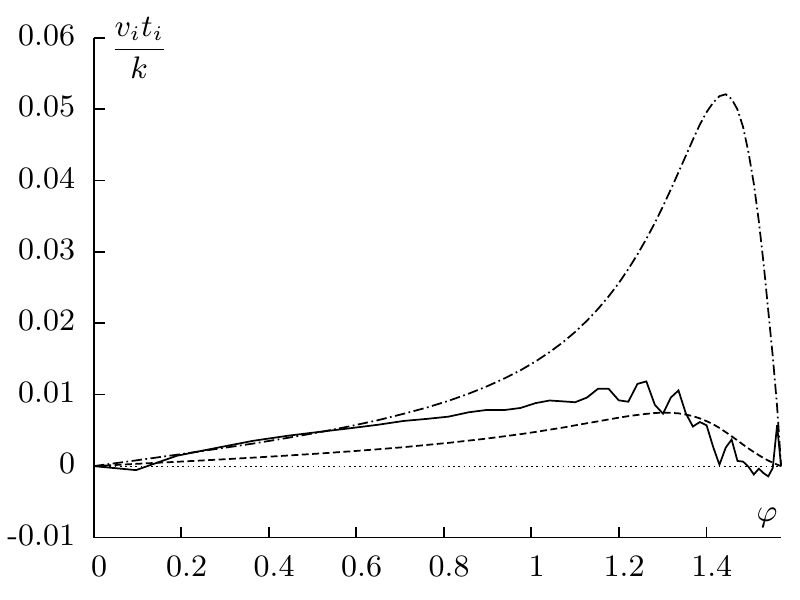}
        \label{fig:profile-vel:inner}
    }\\
    \subfloat[the semi-major axis of the outer cylinder]{
        \includegraphics[width=0.5\textwidth]{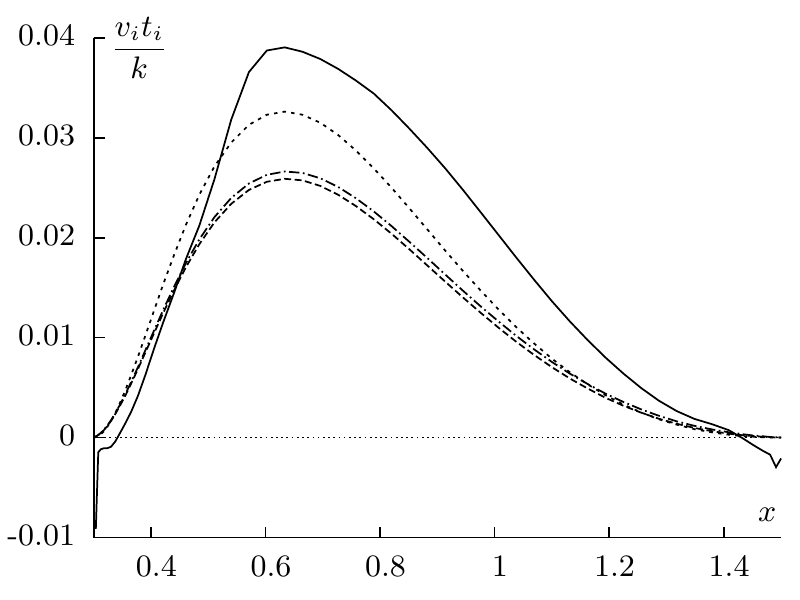}
        \label{fig:profile-vel:bottom}
    }
    \subfloat[the semi-minor axis of the outer cylinder]{
        \includegraphics[width=0.5\textwidth]{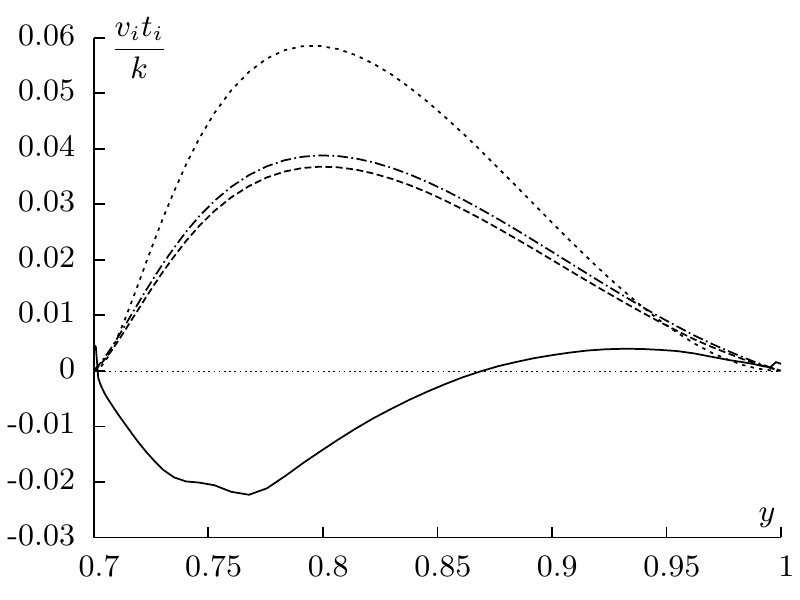}
        \label{fig:profile-vel:left}
    }
    \caption{
        The profile of the boundary tangential velocity. The angle \(\varphi\) corresponds
        to the polar coordinates \(x=r\cos\varphi\), \(y=r\sin\varphi\).
        Unit vector \(t_i\) is directed counterclockwise.
        The following solutions are presented: the Boltzmann equation~\ref{leg:kes},
        the KGF equations with the leading-order~\ref{leg:kgf}, first-order~\ref{leg:first},
        and second-order~\ref{leg:second} boundary conditions.
    }
    \label{fig:profile-vel}
\end{figure}

Consider a gas between two uniformly heated coaxial elliptical cylinders arranged so
that its large axes are rotated through the angle \(\beta=\pi/2\).
\(a_0\) and \(b_0\) are semi-axes of the inner cylinder, \(T_0\) is its temperature,
\(a_1\) and \(b_1\) are semi-axes of the outer cylinder, \(T_1\) is its temperature.
Numerical analysis of the steady-state problem for
\[ a_0 = 0.3, \quad b_0 = 0.7, \quad T_0 = 1, \quad a_1 = 1.5, \quad b_1 = 1, \quad T_1 = 5 \]
is available in the literature.
In particular, a statistical simulation (DSMC) for \(0.1\le\Kn\le5\) is presented in~\cite{Sone1998},
the moment of the forces acting on the cylinder depending on the rotation angle \(\beta\)
is studied on the basis of KGF equations in~\cite{Rogozin2014}.

In the present study, the numerical solution of the Boltzmann equation for the hard-sphere model is compared
with the asymptotic solution based on the KGF equations and boundary conditions of different order.
Due to the symmetry of the problem, consider only the first quadrant (\(x>0\), \(y>0\)),
assuming that the common axis of the cylinders is located in the origin of coordinates.
There are several types of flows depending on the Knudsen number.
In the continuum limit, the nonlinear thermal-stress flow occupies the entire volume
and is directed counterclockwise~\cite{Sone2007,Rogozin2014}.
For \(\Kn>0.1\), on the contrary, the clockwise flow dominates,
while the nonlinear thermal-stress flow occupies only small region most remote from the inner cylinder~\cite{Sone1998}.
The strong near-boundary clockwise flow is driven by the tangential gradient of the gas temperature.
For \(\Kn<0.1\), the dependence of the flow regime on the Knudsen number has not been studied in the literature.
The competition pattern between the first- and second-order nonlinear thermal-stress flows
has also not been obtained until now.
In particular, it requires a huge computational effort to achieve an acceptable signal-to-noise ratio
and discern the structure of these flows by the DSMC method.

For \(\Kn=0.02\), the structured physical mesh consists of \(N_V=2401\) of quadrilateral cells,
which is constructed by the transfinite interpolation method implemented in the GMSH package~\cite{gmsh2009}.
The longitudinal cell edges are nearly tangent to the isothermal surfaces,
and the transverse cell edges to the temperature gradient.
Near the cylindrical surfaces, especially in the region of large temperature gradient,
the physical mesh is refined so that the minimum width of the cell is equal to \(0.046\) mean free paths.
The symmetric nonuniform grid M4 (see Table~\ref{table:velocity_grids}) is constructed in the velocity space.
The distance between nodes increases quadratically along each axis.
This velocity grid provides a sufficiently accurate approximation of the distribution function
for a wide temperature range from \(T_0\) to \(T_1\).

It takes about \(10^5\) iterations with \(5\times 10^4\) cubature points to reach the steady state
of the numerical solution of the Boltzmann equation,
if the initial distribution is determined from the asymptotic solution.
This computation lasts several days on a personal computer with CPU \(4\times3\)~GHz.
The steady-state time averaging helps to reduce a statistical noise
arising from the random shifting of the cyclic Korobov lattice.

The velocity field obtained by different methods is shown in Fig.~\ref{fig:elliptic:flow}.
The second-order boundary conditions (Fig.~\ref{fig:elliptic:flow-asym})
does not change the qualitative picture of the flow in comparison with the leading-order ones,
but only weaken the whole flow due to the velocity jump and second-order thermal slip.
The numerical solution of the Boltzmann equation, however, shows a different flow pattern,
where another clockwise flow competes against the counterclockwise nonlinear thermal-stress one.
This flow occurs in the region where the temperature gradient and curvature of the boundary surface reach their maximum
but outside of the Knudsen layer; therefore, it cannot be described directly through boundary conditions.
Indeed, \(u_{iH1}=0\) on the boundary, because \(t_i\Pder[T_{H0}]{x_i}=0\) (\(t_i\) is a unit vector tangent to the boundary).
The second-order thermal-stress slip is counterclockwise (\(a_4>0\)),
but is balanced by the third-order terms associated with the curvature, proportional to \(\kappa t_i\Pder[T_{H1}]{x_i}\)
(\(\kappa\) is the curvature of the boundary surface).
In other words, \(u_{iH2}t_i\) and \(u_{iH3}t_i\) have opposite signs.
Numerical analysis with the nonlinear second-order boundary conditions has not been carried out,
since the coefficient in front of \((t_i\Pder[T_{H1}]{x_i})(n_j\Pder[T_{H0}]{x_j})\) is unknown.

As can be seen from Fig.~\ref{fig:elliptic:flow},
the leading-order asymptotic solution incorrectly describes the velocity field even for \(\Kn=0.02\).
Indeed, the considered boundary-value problem leads to the region,
where the temperature gradient is comparable with the inverse Knudsen number,
which means \(kn_i\Pder[f]{x_i} = \OO{f}\), but in contrast to the Knudsen layer,
the temperature gradient decreases slowly.
In this region, strictly speaking, the asymptotic solution in the form~\eqref{eq:sum_solutions} cannot be found;
however, it seems the correct flow pattern can be obtained by the use
of the next-order (hitherto unknown) equations for \(T_{H1}\), \(u_{iH2}\), \(p_{H3}\).
In this case, the next-order part of the asymptotic solution is comparable to the leading-order one.

Before proceeding to analysis of the boundary profiles,
specify the method of calculating the temperature on a diffuse-reflection surface.
Actually, the finite-volume solution provides solution only in the cell centers.
Since the temperature has a weak logarithmic singularity in the Knudsen layer,
the boundary temperature is extrapolated as \(Ay\ln{y}+B\),
where \(A\) and \(B\) are constants, \(y\) is the distance from the boundary.
In the previous problem, it was sufficient to use the linear extrapolation,
because the width of the boundary cell and normal temperature gradient were less.

Fig.~\ref{fig:profile-temp} and~\ref{fig:profile-vel} show the temperature and velocity profiles on the boundary surfaces.
The inclusion of the first-order temperature jump in the boundary conditions
yields a significant improvement to the asymptotic temperature field,
while the second-order temperature jump gives only a small correction.
This is due to the fact that \(n_i\Pder[T_{H0}]{x_i}\) and \(n_in_j\Pderder[T_{H0}]{x_i}{x_j}\)
are comparable but much larger than \(T_{H0}\).
In Fig.~\ref{fig:profile-temp}, due to the cubature error for temperature,
the numerical solution exceeds the asymptotic one less than \(0.004\);
however, in Fig.~\ref{fig:profile-temp:left} and for \(\varphi>\pi/3\) in Fig.~\ref{fig:profile-temp:inner},
the difference between the solutions is larger due to the considerable discrepancy between the velocity fields \(u_i/k\),
affecting the temperature field through the energy equation~\eqref{eq:asymptotic1_T}.
Indeed, Fig.~\ref{fig:profile-vel:left} shows that the corresponding values of \((u_i/kT)\Pder[T]{x_i}\) differ even in sign.

The second-order boundary conditions provide a better approximation of the numerical solution
at the outer cylinder (Fig.~\ref{fig:profile-vel:outer}), but do not at the inner one (Fig.~\ref{fig:profile-vel:inner}).
As mentioned above, this is due to the fact that the boundary conditions for \(u_{iH3}\) are \(\OO{u_{iH2}/k}\)
in the region of maximum temperature gradient, but are not included in the solution of the KGF equations.
Sharp fluctuations in the numerical solution of the Boltzmann equation (especially in Fig.~\ref{fig:profile-vel:inner})
come from the discretization error in the velocity space, but do not exceed \(10^{-4}\) in absolute value of \(u_i\).

\section{Conclusion}

Some numerical examples of slow nonisothermal slightly rarefied gas flows,
as well as their influence on the temperature field, have been analyzed.
With the help of the Tcheremissine's projection-interpolation discrete-velocity method,
the flow patterns have been obtained with an accuracy that is unattainable by the conventional DSMC method.
Adjusting the discretization of the velocity space is the main difficulty for nonlinear problems,
including flows driven by large temperature variations.
In the present work, this problem has been handled by using nonuniform velocity grids,
which, however, complicates the calculations.
In particular, the conservative projection onto a nonuniform grid requires
a more complex algorithm and more cubature points.
In addition, the cubature error for distributions that are close to the Maxwellian, typical for slow flows, grows significantly.
Nevertheless, for small Knudsen numbers, the obtained numerical solutions coincide with the asymptotic ones with high accuracy.
Hence, projection-interpolation method on nonuniform grids can be recommended
for simulation a wide range of slow nonisothermal gas flows.

Upon comparative analysis of the numerical and asymptotic solutions of the Boltzmann equation for a hard-sphere gas,
the Kogan--Galkin--Friedlander equations with appropriate boundary conditions have been shown to correctly describe
slow nonisothermal gas flows for sufficiently small Knudsen numbers.
If there is a temperature variation comparable to unity on the scale of the mean free path,
then the fluid-dynamic approach is inapplicable to describe the rarefied gas.
Instead, the problem should be discussed within the framework of the Boltzmann equation.

The boundary conditions that comprise the first-order temperature and velocity jumps along with the leading-order thermal creep
have been shown to significantly improve the accuracy of an asymptotic solution,
since large normal temperature gradients are typical for the considered problems.
Including the next-order terms in the boundary conditions for the KGF equations gives a small correction
and hence can be recommended only for high-accuracy numerical analysis.

The slip/jump coefficients in the boundary conditions, together with the Knudsen-layer functions,
are obtained from the numerical solutions of the corresponding one-dimensional linearized problems.
Currently, they are known only for the hard-sphere molecular potential and some model equations~\cite{Takata2016}.
This fact limits the use in applied problems.
Furthermore, the corresponding second-order Knudsen-layer problems for the nonlinear boundary-condition terms,
which are reduced to the inhomogeneous linearized Boltzmann equation,
are not analyzed even for a hard-sphere gas.

\section*{Acknowledgement}

The author is grateful to Oscar Friedlander for fruitful discussions, which served as the starting point to the present study,
to Vladlen Galkin for his detailed and valuable comments that helped to improve the presentation of the material,
and to Felix Tcheremissine for his encouragement and interest in this work.

\bibliography{manuscript}
\bibliographystyle{elsarticle-num}

\end{document}